\documentclass[fleqn,usenatbib]{mnras}

\usepackage{newtxtext,newtxmath}

\usepackage[T1]{fontenc}
\usepackage{ae,aecompl}

\usepackage{graphicx}	        
\usepackage{amsmath}	        
\usepackage{amssymb}	        
\usepackage{gensymb}            
\usepackage{multirow}           
\usepackage{ulem}               
\usepackage[dvipsnames]{xcolor} 
\usepackage{enumitem}           

\newcommand{\code}[1]{\texttt{#1}}

\newcommand\numberthis{\addtocounter{equation}{1}\tag{\theequation}}

\title[CLUES about M33]{CLUES about M33: the reversed radial stellar age gradient in the outskirts of Triangulum galaxy}

\author[Mostoghiu et. al]
       {Robert Mostoghiu$^{1,2}$\thanks{robert.mostoghiu@uam.es},  Arianna Di Cintio$^{2,3,4}$\thanks{adicintio@iac.es}, Alexander Knebe$^{1,5,6}$, 
\vspace{-.3cm}\newauthor{Noam I. Libeskind$^{2}$, Ivan Minchev$^{2}$ \& Chris Brook$^{3,4}$} \\
\vspace{-.3cm}
$^{1}$Departamento de F\'isica Te\'{o}rica, M\'{o}dulo 15, Facultad de Ciencias, Universidad Aut\'{o}noma de Madrid, E-28049 Madrid, Spain\\
\vspace{-.3cm}
$^{2}$Leibniz-Institut f\"{u}r Astrophysik Potsdam (AIP), An der Sternwarte 16, D-144 Potsdam, Germany\\
\vspace{-.3cm}
$^{3}$Instituto de Astrof\'{i}sica de Canarias, Calle Via L\'{a}ctea s/n, E-38206 La Laguna, Tenerife, Spain\\
\vspace{-.3cm}
$^{4}$Universidad de La Laguna. Avda. Astrof\'{i}sico Fco. S\'{a}nchez, La Laguna, Tenerife, Spain\\
\vspace{-.3cm}
$^{5}$Centro de Investigaci\'{o}n Avanzada en F\'isica Fundamental (CIAFF), Facultad de Ciencias, Universidad Aut\'{o}noma de Madrid,\\ \vspace{-.3cm} 28049 Madrid, Spain\\
\vspace{-.3cm}
$^{6}$International Centre for Radio Astronomy Research, University of Western Australia, 35 Stirling Highway, Crawley,\\  \vspace{-.3cm} Western Australia 6009, Australia
}

\date{Accepted XXX. Received YYY; in original form ZZZ}

\pubyear{2018}
\begin{document}
\label{firstpage}
\pagerange{\pageref{firstpage}--\pageref{lastpage}}
\maketitle

\begin{abstract}
HST/ACS observations along the major axis of M33  show that the  mean age of its stars decreases with increasing distance from the galaxy center. Such a behavior is consistent  with an inside-out growth of the disc. However, in the outermost observed field, at $r\simeq$11.6 kpc, a  reversal of  this  gradient is detected, with old stars found in high percentages beyond this radius. In this work we investigate the origin of such a  reversal in  stellar  age gradient, by using  a simulated M33 analogue from the Constrained Local UniversE Simulations (CLUES). The simulated M33 is similar to the observed one in terms of  mass, rotation velocity, surface brightness and, similar to what has been reported in observations, shows a stellar age turnaround at large radii.  We demonstrate that this reversal is  mostly a result of stellar accretion from old satellite galaxies and, to a lesser extent, of stellar migration of in-situ stars.  The old  accreted stars, with formation times t$_{f}<4$ Gyrs,  are kinematically  hot and can be differentiated from the in-situ stars by their high velocity dispersion and the fact that they do not have rotationally-supported orbits. In the future, obtaining kinematic information of the  stars in the outskirt of M33 will help to verify this scenario.

\end{abstract}




\section{Introduction} \label{sec:intro}
In a $\Lambda$CDM universe, spiral galaxies consist of a disc component made of stars, cold gas and dust, a central bulge and a stellar halo, all embedded in a dark matter halo \citep{white78}. The disc component can be separated into two different parts: the thin disc, and the thick disc \citep{Burstein79, Gilmore83}. These two components are defined by examining the vertical scale height of stars when separated by age (e.g. \citealt{Haywood13,Bensby14}) or metallicity (e.g. \citealt{Fuhrmann08,Bensby14}). The stars in the thin disc component are formed by gas accretion at the later stages of galaxy formation and they have a wide range of ages \citep{Yoachim06}. The stars in the thick disc, however, are older and  their origin is still debated (e.g. \citealt{Brook04, Villalobos08, Minchev15}).

The distribution of stars in galactic discs is also an ongoing research area. One of the favorite modes for the mass assembly of a galaxy is the ``inside-out'' scenario \citep{Chiappini97, Mo98, Brook12, Pilkington12, Bird13}. In the inside-out growth proposal, the inner disc is thought to assemble first as a consequence of the high density of accreted gas residing in the center of the galaxy's potential well. Thus, the fraction of young stars is expected to increase with galactocentric radius. Several galaxies have been found to be compatible with such a growth model \citep{Perez13, SanchezBlazquez14, Tacchella15}.

Recent observations regarding the ages of stars in the neighboring galaxy M33 indicate that this galaxy is compatible with an inside-out disc growth scenario, in which old stars are detected in the inner region of the galaxy, while young, disc stars tend to naturally be found in the outskirts of the disc \citep{Williams09, Barker11}. Specifically, these observations made use of the \textit{Hubble Space Telescope Advance Camera for Surveys} (HST/ACS), to derive the cumulative star formation history (SFH) along M33's major axis and for different radii. The SFH was derived using the synthetic color-magnitude diagram (CMD) fitting method. CMDs were obtained by measuring resolved stellar photometry using the ACS module of the \code{DOLPHOT} software package \citep{Dolphin00}. Assuming an initial mass function and stellar evolution isochrones, a fitting is performed on the CMD to obtain the star formation rate at their respective ages and metallicities. \citet{Williams09} and \citet{Barker11} showed that within $\approx 9$ kpc from M33's center, the mean age of stars decreases as one moves further out from the galactic center.  They also showed that at radii greater than $\approx 9$ kpc, however, this \textit{age gradient} reverses, such that the mean age of stars increases as one approaches the outer region of M33. The age gradient thus reverses from decreasing mean stellar age with radius (within 9kpc) to increasing mean stellar age with radius (beyond 9kpc). Note that the age gradient reversal is accompanied by a surface brightness and stellar mass surface density break beyond 8 kpc \citep{Ferguson07,Barker11}, whose physics remains contentious (see \citealt{trl17}  for a recent review of the subject using simulations). 

Similar age profiles have been seen in both simulations (e.g. \citealt{Roskar08a,Roskar08b, SanchezBlazquez09,MartinezSerrano09,RuizLara16a}) and observations of disc galaxies \citep[e.g.,][]{Bakos08, Yoachim12, zheng15,RuizLara16b}, yet the origin of the reversal is not clear. Several explanations for the reversal have been proposed: stellar migration, in which the inner disc forms inside-out and the region beyond the upturn radius is populated with stars that migrated from the inner disc  \citep{Roskar08a,Roskar08b,RuizLara16a}; projection effects, that cause a contamination and overlap of stars from different galactic components \citep{Barker11}; a decrease in the gas volume density in the disc, which induces a break in the star formation density which itself coincides with the radius where the gas disc begins to warp \citep{SanchezBlazquez09}; or old stars coming from mergers that, due to their significant energy, remain in orbits at large, outer radii \citep{Gill05,Sales2007,Brook12,RuizLara16a}.
 
 In this paper we explore the age gradient of a simulated M33 analogue galaxy, formed in a constrained Local Group environment as part of the CLUES\footnote{\url{www.clues-project.org}} project \citep{Gottloeber10,Carlesi16}.  The initial conditions have been constrained by observational data such that the z=0 cosmography is forced to reproduce the real local environment \citep{Libeskind15,Sorce16}. The simulated M33 analogue shares many properties with the observed M33 and was formed in a similar environment. This means that our analysis of the origin of the M33 analogue may provide insights into the mechanisms driving the age gradient, in the real M33, in particular the reversal of the age gradient that is observed. 
 
The paper is organized as follows. In Section \ref{sec:m33_counterpart} we present the simulated M33. In Section \ref{sec:m33_counterpart.subsec:construct_numerical_M33} we present the simulation's properties. In Section \ref{sec:m33_counterpart.subsec:validate_numerical_M33} we focus on the features of our candidate galaxy. The reversal of the age gradient in the SFH of M33 is presented in Section \ref{sec:m33_age_grad}. In Section \ref{sec:m33_age_grad.subsec:presentation} we present the adopted methods to  analyze  the age reversal, in Section \ref{sec:m33_age_grad.subsec:explanation} we discuss the implications of our study, and in Section \ref{sec:m33_age_grad.subsec:observation_predict} we give observational predictions. Finally, in Section \ref{sec:conclusions} we summarize our results.

\begin{figure*}
    \hspace{-0.5cm}
    \includegraphics[width=1.\textwidth]{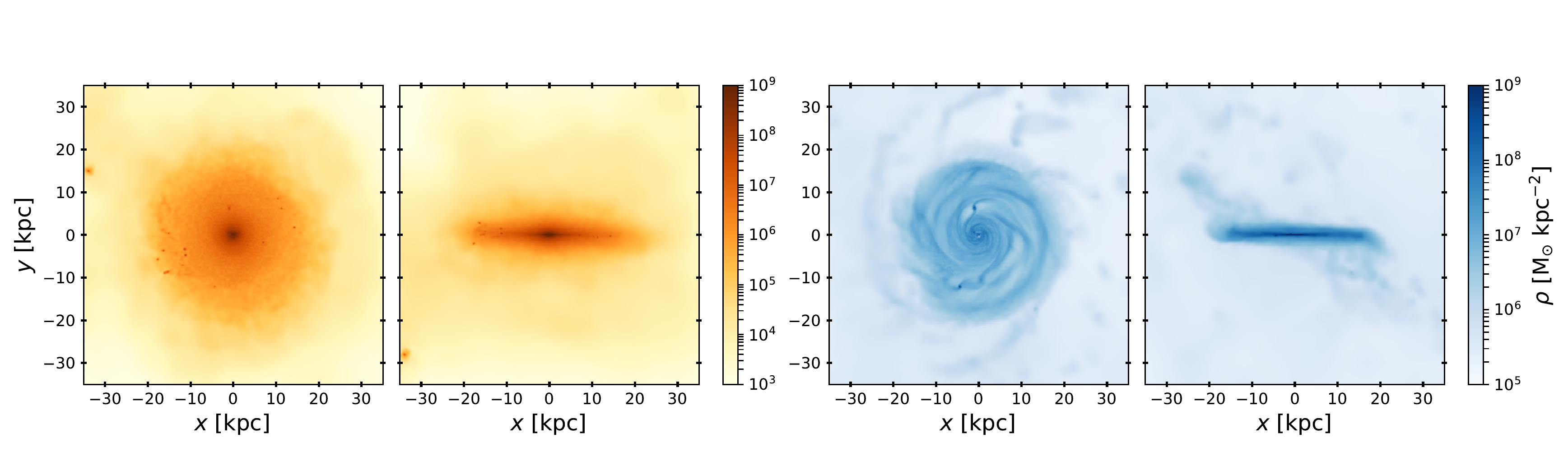}
    \caption{Stellar and gas mass densities of the simulated M33 galaxy at z=0. \textit{Left}: stellar mass density, face-on and edge-on. \textit{Right}: gas density, face-on and edge-on. Spiral features and a thin warped disc of gas  are visible, in agreement with observational data (e.g. \citealt{Corbelli1997,Kam17}).}
    \label{fig:density_gasANDstars}
\end{figure*}

\section{A Simulated Counterpart of M33} \label{sec:m33_counterpart}
In this section we describe the properties of the M33 candidate found within one of the hydrodynamical numerical simulations performed as parta of the \textit{Constrained Local UniversE Simulations} (CLUES) project \citep{Gottloeber10}.

\subsection{Constructing the numerical M33} \label{sec:m33_counterpart.subsec:construct_numerical_M33}
In order to create constrained simulations of the Local Group (LG), the initial conditions are tuned using as observational constraints peculiar velocities obtained from the MARK III catalog \citep{Willick1997}, surface brightness fluctuations \citep{Tonry01}, local volume galaxy catalogs \citep{Karachentsev04}, and the position and virial properties of nearby X-ray selected clusters of galaxies \citep{Reiprich02}. Then, using the Hoffman-Ribak algorithm \citep{Hoffman1991} and the observational constraints, the initial conditions are generated as constrained realizations of Gaussian random fields on a $256^3$ mesh.
Note that only large, linear scales are constrained: since they are non-linear, properties of the LG  such as mass, relative positions and velocities, are unconstrainable. Therefore, low resolution initial conditions are run to $z=0$, and those simulations which reproduce a structure resembling the LG are selected for high resolution resimulation. Numerical power is added following a zoom-in technique \citep{Katz94,Navarro94}. The end result is a local group, selected in a frequentist manner, embedded in the proper constrained large scale structure. In the current work, the simulated LG includes a third galaxy which resembles M33 in terms of mass and placement, being at $\simeq 800$ kpc from the other two main halos.

The simulation was run using the standard $\Lambda$CDM \textit{Wilkinson Microwave Anisotropy Probe} 3 (WMAP3) cosmology \citep{Spergel07}, with $\Omega_{M}= 0.24$, $\Omega_{b}=0.042$, $\Omega_{\Lambda}=0.76$, $\sigma_{8}=0.75$, $h=0.732$, and an $n=0.95$ slope of the power spectrum. We used the parallel TreeSPH code \code{GASOLINE} (for further details  see \citet{gasoline,wadsley04} and references therein) to simulate a cosmological box with side length $L_{\rm box}=64h^ {-1}$ Mpc, and effective particle resolution of $m_{\rm DM}=2.1\cdot 10^{5}h^{-1}$M$_{\odot}$ and $m_{\rm gas} = 4.4\cdot 10^{4}h^{-1}$M$_{\odot}$. 

These hydrodynamical simulations have been presented and used to explore dwarf galaxies in the Local Volume \citep[see][where full details are found]{santos16,santos17}.  In particular, the simulated galaxies were shown to match a range of scaling relations, including the relations between stellar and halos mass, stellar and HI gas mass, size and stellar mass, and the Tully-Fisher relation; meaning we can have a degree of confidence in our analysis.  The simulations include a UV background, gas cooling, and star formation, with the stars feeding energy back into the interstellar medium (ISM) gas. Gas is eligible to form stars when it reaches temperatures below 15000 K in a dense environment, with minimum density threshold of 10 amu/cm$^3$. Blastwave supernova feedback is included \`{a} la \cite{Stinson06}, allowing an efficient regulation of star formation within galaxies. The stellar particles are formed with an initial mass of $m_{\star}=1.5 \cdot 10^{4}h^{-1}$M$_{\odot}$.  

To identify halos in the simulation we used the MPI+OpenMP hybrid halo finder \textit{AMIGA Halo Finder} (\code{AHF})\footnote{\url{http://popia.ft.uam.es/AHF}} \citep{Gill04a, Knollmann09}, which locates local overdensities in an adaptively smoothed density field as potential halo centers and automatically identifies halos, subhalos, subsubhalos, etc. For every found halo, \code{AHF} calculates its virial radius $r_{\rm vir}$ as the radius $r$ at which the density $\rho(r)=M(<r)/(4\pi r^{3}/3)$ drops below $\Delta_{\rm vir}\rho_{\rm b}$, where $\Delta_{\rm vir}$ is a cosmological model and time dependent threshold parameter, and $\rho_{b}$ is the cosmological background matter density. The threshold $\Delta_{\rm vir}$ is computed using the spherical top-hat collapse model. For the cosmology that we are using, $\Delta_{\rm vir}$ = 355 at $z = 0$  \citep{Bryan98}.

To trace halos through the snapshots we build merger trees with \code{MergerTree}, a tool that comes with \code{AHF}. \code{MergerTree} identifies counterpart objects in the same simulation at different redshifts. \code{MergerTree} follows each halo identified at redshift z = 0 backwards in time, identifying as the main progenitor at some other redshift the halo that both shares the most particles with the present halo and is closest in mass. More details can be found in \citet{Srisawat13}.

Finally, for the analysis of the identified halos we used the Python-based package \code{PYNBODY}\footnote{\url{https://github.com/pynbody/pynbody}} \citep{pynbody}.

\subsection{Validating the numerical M33} \label{sec:m33_counterpart.subsec:validate_numerical_M33}

In this sub-section we focus our attention on the M33 candidate found in the Local Group simulation, examining how the properties of the candidate compare to the observed one.
We start by showing, in Fig. \ref{fig:density_gasANDstars}, a visualization of the simulated galaxy at z=0, with the face-on and side-on views of the stellar and gas density of our M33 candidate. 
A warped disc component can be seen in the edge-on view, similarly to the warped disc reported in observations of M33 (e.g. \citealt{Corbelli1997, Kam17}).
Moreover, similar to the observed M33, our simulated counterpart shows well defined spiral features, clearly visible in the face-on gas density plot.

In order to compare more quantitatively with observational results, we present in Tab.\ref{tab:clues_halo_mass} the virial and stellar mass of the M33 candidate, and we compare these values with the observational M33 data reported in \citet{Corbelli14} and \citet{Kam17}.
For the simulation, the virial mass is computed as the total mass within the virial radius at $z=0$, while the stellar mass is the sum of all the star particles found within the galaxy and its halo.
We note that with a value of $M_{\rm vir}=2.7\cdot10^{11}M_{\odot}$ and $M_{\star}=5.1\cdot10^{9}M_{\odot}$, our simulated M33 lies on the expected $M_{\star}$-$M_{\rm halo}$ relation, or abundance matching relation (e.g. \citealt{Moster10}).

\begin{table}
\centering
\caption{Total virial mass and  stellar mass of simulated and observed M33. Observational data from \citet{Corbelli14,Kam17}, see text for more details.
}
\label{tab:clues_halo_mass} 
\begin{tabular}{cccc}
\hline
\hline
M33 & $M_{\rm vir}$($10^{11}$M$_{\odot}$)& $M_{\star}$($10^{9}$M$_{\odot}$)&$M_{\rm HI}$($10^9$M$_{\odot}$)\\
\hline
SIM& 2.7           & 5.1         & 2.8  \\ 
\hline
OBS& 4.3$\pm$1.0   & 4.8$\pm$0.6 & 1.9 $\pm$0.4 \\ 
\hline
\hline
\end{tabular}
\end{table}

For the observed Triangulum galaxy, the virial mass and the stellar mass, together with the concentration of dark matter halo, are obtained by considering the composite probability of three events: the dynamical fit to the measured rotation curve of M33, the stellar mass determined via synthesis models, and the concentration-mass relation $c(M)$ found in numerical simulations of structure formation for a $\Lambda CDM$ cosmology (see \citealt{Corbelli14} for more details).
Tab.\ref{tab:clues_halo_mass} indicates that our simulated M33 is very similar, in both total and stellar mass, to the observed M33, justifying further comparisons.

\begin{figure}
	\hspace{-.6cm}\includegraphics[width=1.1\columnwidth]{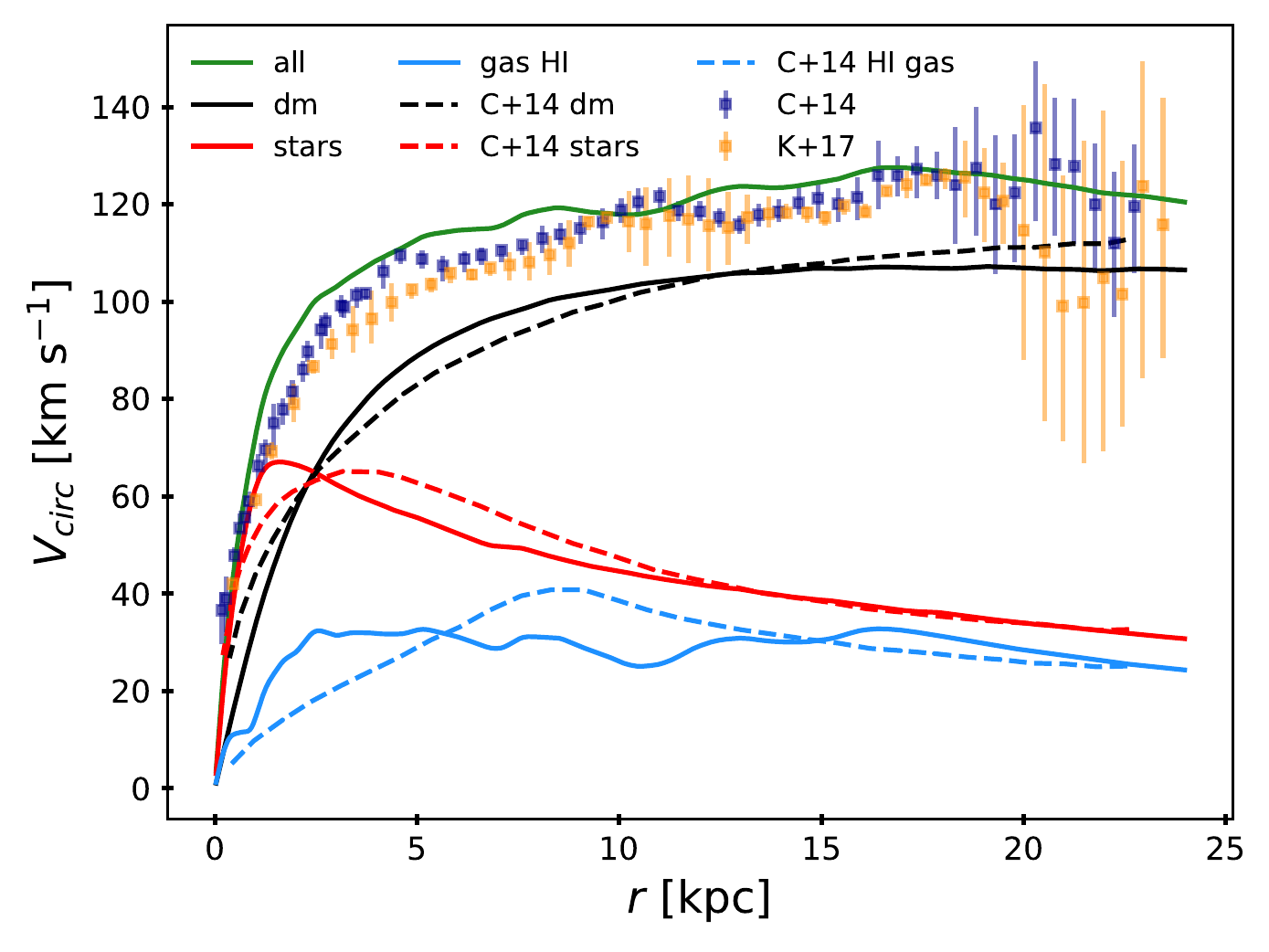}
    \caption{Circular velocity profiles of DM halo (black), stars (red) and  HI gas (light blue) of the simulated M33 galaxy (solid lines) compared with observations from \citet{Corbelli14} (dashed lines).
 The total velocity profile is shown in green for the simulated M33 and as blue and yellow points with error bars for the  \citet{Corbelli14} and  \citet{Kam17} data, respectively.
 The observed and simulated rotation curves agree within a few percents at all radii. See the online version for details.}
    \label{fig:circ_vel_prof_corb14}
\end{figure}

In Fig.\ref{fig:circ_vel_prof_corb14} we study the mass component distribution of the galaxy, showing the circular velocity profile of  dark matter (black lines), HI gas (light blue lines), and stars (red lines), alongside observations from \citet{Corbelli14} and \citet{Kam17}.  Observations are shown as dashed lines and simulations, as solid ones. The total velocity is shown as a green solid line for  the simulated M33 and as points with error bars for the observed one.
The simulated velocity profiles are computed by using the gravitational potential of the particles in the galactic midplane after placing the galaxy face-on using its total angular momentum. The HI gas component was obtained directly from the simulations, which solves the Saha equation to calculate the ionization state based on the pressure and temperature.

The total  velocity curve matches the observed one quite well, reaching a maximum of $V=127.6$  km/s at a radius of $r=16.9$ kpc, like the real M33.
However, some of the  individual components have a rotation curve that rises faster  in the innermost region of the galaxy, compared to the profiles derived from observational data in \citet{Corbelli14}.
The HI gas component is closer to the observational data in the outer region of the galaxy ($r\gtrapprox13$ kpc), while in the inner region it underestimates the peak velocity. Further, the slightly higher contribution from stars at the center of the simulated galaxy reflects the contribution of its larger-than-observed bulge (see below). Nevertheless, the overall agreement between the total simulated rotation curve and the observed one is quite good, with differences of a few per cent at most, depending on the radii.

Our last comparison to observations will now go into even more details by mimicking observations as closely as possible. We therefore set the simulated M33 in a configuration which resembles the observed one. Starting from an initial face-on view, where the disc lies on the $xy$ plane, we inclined the M33 candidate $60$ degrees around the $x-$axis to reach the reported inclination of M33 to the line of sight from the Earth, $i\approx50^{\circ}-60^{\circ}$ (e.g. \citealt{Patterson1940,Verley09}). Furthermore, since observational data has been measured along the major axis of the galaxy to avoid stellar contamination from different galactic components, we select the major axis of the inclined simulated M33 in a similar fashion by applying a position cutoff in the axis perpendicular to the inclination axis, i.e. we select star particles with $|y|\leq5$ kpc. Note that, since our galaxy is not circularly symmetric, there is a degeneracy in the initial face-on view of the galaxy; different initial face-on views (as generated by rotations about our $z$-axis) produce different profiles of projected quantities after inclining the galaxy. For our analysis we selected a face-on view which best reproduces the observed surface brightness profile\footnote{We studied 36 initial $z-$axis rotations of the initial face-on view, spanning 360 degrees, and found that $\sim 64$ per cent of the  initial configurations present a strong break in the profile, while the rest show weak-to-null breaks after the $60$ degree rotation.}, to be discussed now.

\begin{figure}
	\includegraphics[width=1\columnwidth]{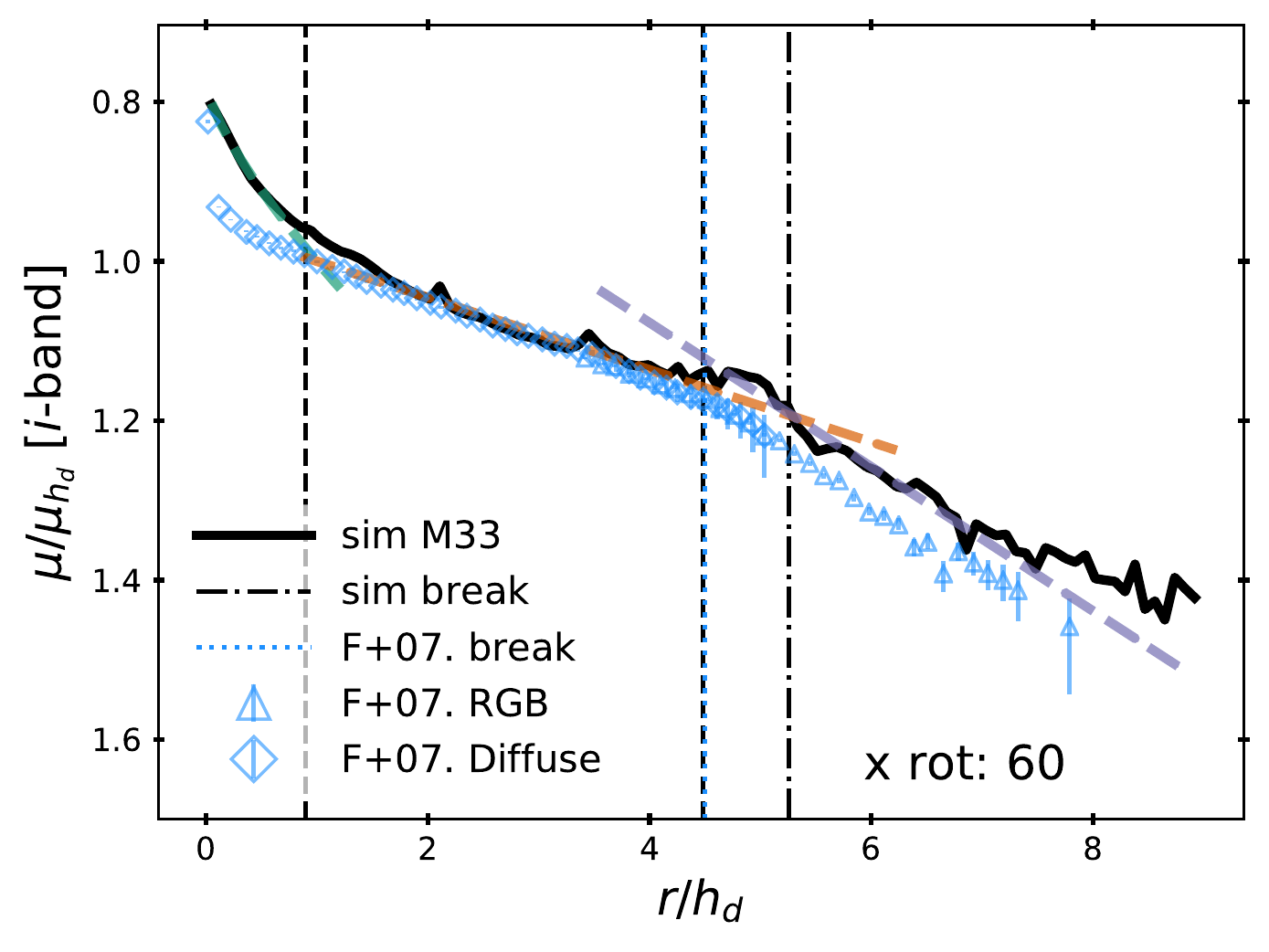}
    \caption{Surface brightness of the simulated M33 for the inclined configuration, normalized by the value at the inner disc's scale radius $h_{d}$. The best-fit values of the $3-$component fit (colored dashed lines) are presented in Tab. \ref{tab:best_fit_surfb}. The vertical black dashed lines show the $r=3$ kpc and $r=15$ kpc regions. Observational values from \citet{Ferguson07} are represented as blue markers, with its corresponding break at $r=4.5h^{M33}_{d}=8.1$ kpc (blue dotted line, on top of the $r=4.5h_d=15$ kpc dashed line). Similar to what is observed in the real M33, a break can be seen in the simulated galaxy at a similar position, $r\sim 5.3h_{d}=17.6$ kpc (black dash-dotted line), as well as a truncation at $r\sim 7.6h_{d}=25$ kpc.}
    \label{fig:surfb_fit_withObs}
\end{figure}

Once the simulated M33 is in its inclined configuration, we performed a bulge/disc decomposition in order to compute the galaxy's disc scale length (for a detailed description of the analysis, see Appx. \ref{app:disc_scale_length}). We fit the $i-$band surface brightness profile of the galaxy with a $3-$component model, i.e. a combined inner and outer exponential discs plus a Sersic bulge, to account for breaks in the profile. We obtain a  inner disc scale-length of $h_{d}= 3.3\pm0.1$ kpc. The observed M33 has a somewhat shorter scale length for the disc, $h^{M33}_d=1.8$ kpc \citep{Verley07,Ferguson07,Verley09,Corbelli14}. To account for this difference and in order to compare our simulated profiles with the observational data, respectively, we normalize the $x$-axis to the respective inner exponential disc scale length and the $y$-axis to the respective value at that position. In Fig. \ref{fig:surfb_fit_withObs} we show the (normalized) surface brightness profile. Along the simulated profile, we present the observed $i-$band surface brightness profile from \citet{Ferguson07}. We find the best match to observational data within the region $0.9<r/h_{d}<4.5$ (or in physical units $3<r<15$ kpc) -- as indicated by the two vertical dashed lines. In the inner region of our numerical M33 ($r\lessapprox3$ kpc, left dashed line), we can identify an excess of light from the bulge component. From $r\gtrapprox15$ kpc onwards (as indicated by the right dashed line), the M33 candidate follows closely the observed light distribution. Additionally, the simulated profile features a down-bending disc break at $r\sim 5.3h_{d}=17.6$ kpc, which correlates with the radius at which the age reversal is found (see Sec. \ref{sec:m33_age_grad}). The disc break in the observational profile is found at a similar (although slightly smaller) position, at $r=4.5h^{M33}_{d}=8.1$ kpc. Following \citet{Martin-Navarro12,Martin-Navarro14}, at $r\sim7.6h_{\rm d}=25$ kpc we can also identify an up-bending in the surface brightness profile associated with the stellar halo component  of the simulated galaxy, coexisting with a truncation, i.e. a sharp decline in the radial light profile, that allows the stellar halo component to outshine the disc's light distribution. We note that we obtain similar results in the stellar surface mass density profile of the simulated galaxy (not shown here): the $3-$component fit leads to a comparable inner disc scale length, and  we identify a break and a truncation at the same radii. The fact that we observe a break in the both profiles suggests that the reversal in the age gradient is affected by the combined effect of both the radial distribution of the star particles and their ages and metallicities.

We have just demonstrated that our numerical M33 is in fact an adequate replica of the observed M33: it is situated in the right environment, has about the correct size and mass, features comparable kinematics, and even agrees fairly well when observed from the right angle. And as we will see in the next Section, it also features the observed reversal of the age gradient. This motivated us to seek its origin using the constrained simulation of the Local Group and its constituent M33. However, we like to close with the cautionary note that even though our numerical M33 appears to be a reasonable counterpart of the observed one, we need to remind the reader that it is by far a facsimile. As mentioned before, the very nature of following non-linear cosmic structure by means of numerical simulation only allows constraining scales beyond the size of the Local Group; random fluctuations will always enter scales smaller than that. For an elaborate discussion of such effects and restrictions we like to refer to the work of \citet{Carlesi16} where the `Local Group Factory', i.e. a framework for simulating the `near field', has been presented.

\section{The Reversed Stellar Age Gradient in M33} \label{sec:m33_age_grad}
In the previous section we examined some specific properties of our M33 candidate, and we showed that, despite some unavoidable differences we are able to reproduce fairly well the morphology, luminosity, mass and velocity curve of the observed M33 galaxy. But one of the strongest arguments for studying the numerical M33  in our simulated LG is that we found a similar reversal of the age gradient in the cumulative normalized SFH of the simulated M33, as observations report (e.g. \citealt{Williams09,Barker11}). In the next sections we present our  analysis of this phenomenon.

\subsection{Presentation of the Radial Stellar Age Gradient} \label{sec:m33_age_grad.subsec:presentation}

\begin{figure}
\hspace{-.6cm}\includegraphics[width=1.08\columnwidth]{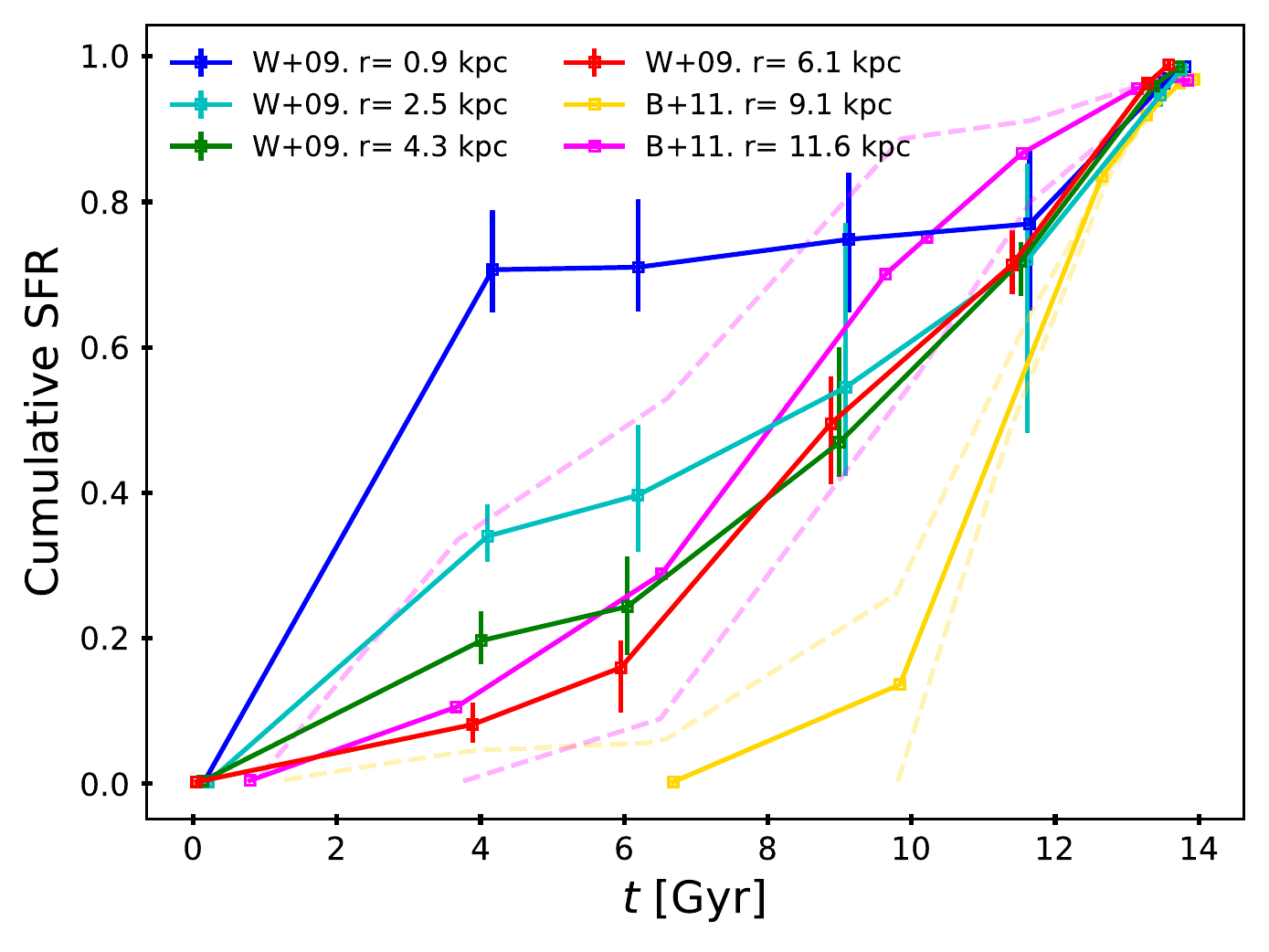}
    \caption{Cumulative normalized SFH of the M33 galaxy measured along its major axis. Data retrieved from  \citet{Barker11} with results from \citet{Williams09}, presented in a revised form for direct comparison with our simulation. The dashed lines represent the 1$\sigma$ contours for the last two radial bins. As we move further out from the galactic center the young stellar  population increases, compatibly with an inside-out growth of the disc. Between $r=9.1$ kpc and $r=11.6$ kpc, however, a significant shift is found to older age (smaller $t$), signifying a reversal in the galactic radial age gradient, that is, the stellar radial age gradient of the galaxy reverses. See the online version for details.}
    \label{fig:obs_sfh}
\end{figure}

In Fig. \ref{fig:obs_sfh} we show the combined observational results of the  cumulative normalized SFH of M33, as reported in \citet{Williams09} and \citet{Barker11}, referred to as W09+B11 from now on. The observations were made along the major axis of M33 using the HST/ACS in the Wide Field Channel, with field of view of $202''x202''$, at radii of $r=0.9$, $2.5$, $4.3$, $6.1$ kpc in \citet{Williams09}, and later extended to $r=9.1$, $11.6$ kpc by \citet{Barker11}. As we move further out from the galactic center the intermediate-to-old star population fraction decreases while  young stars start contributing more and more to the overall budget in the outer disc, compatible with an inside-out growth. Within 0.9 kpc from the galactic center, more than $70$ per cent of stars are old,  having formed in the first 4 Gyrs of the  galaxy assembly, while  less than $10$ per cent of the stars found at  a radius of 6.1 kpc are old. At 9.1 kpc, more than $80$ per cent of the stars are young, specifically younger than 4 Gyrs.

However, as already noted in \citet{Barker11}, once the radius  $r=11.6$ kpc is reached (in magenta), the curve shifts to lower $t$ with respect to $r=9.1$ kpc (in yellow), i.e. the fraction of old stars increases again. Indeed, at 11.6 kpc stars are found  that are  older than the oldest population at $r=9.1$ kpc, even after considering the $1\sigma$ error (dashed lines).
Such an age gradient reversal, with increasing mean stellar age with radius, thus happens for radii beyond 9 kpc.

In order to compare our galaxy to observations we analyze the star particles in our halo in a similar fashion to the measurements  done in W09+B11. After positioning the simulated galaxy face-on, we choose a series of concentric annuli of thickness $\delta_{\rm ann} = 0.8$ kpc, similar to the field of view of the camera used in W09+B11, centered at the same radii of  Fig. \ref{fig:obs_sfh}, $r_{\rm obs}$, for which observations exist.
The annuli's radii will then be $r_{\rm ann} = r_{\rm obs} \pm \delta_{\rm ann}/2$, where $r_{\rm obs}=0.9$, 2.5, 4.3, 6.1, 9.1, 11.6 kpc. We further extend the sampling region up to $r_{\rm ann}=30$ kpc, in order to cover even the outer part of the disc, for which observations of the M33's SFH currently do not exist.
We choose annular regions in order to avoid a definition of a major axis and to get a statistically meaningful result. In order to check that our results are insensitive to \--- and not driven by \--- the inclination of the galaxy, we repeat the same analysis for our (best) inclined configuration described in Sec.~\ref{sec:m33_counterpart.subsec:validate_numerical_M33}.

\begin{figure}
	\hspace{-.6cm}\includegraphics[width=9.8cm]{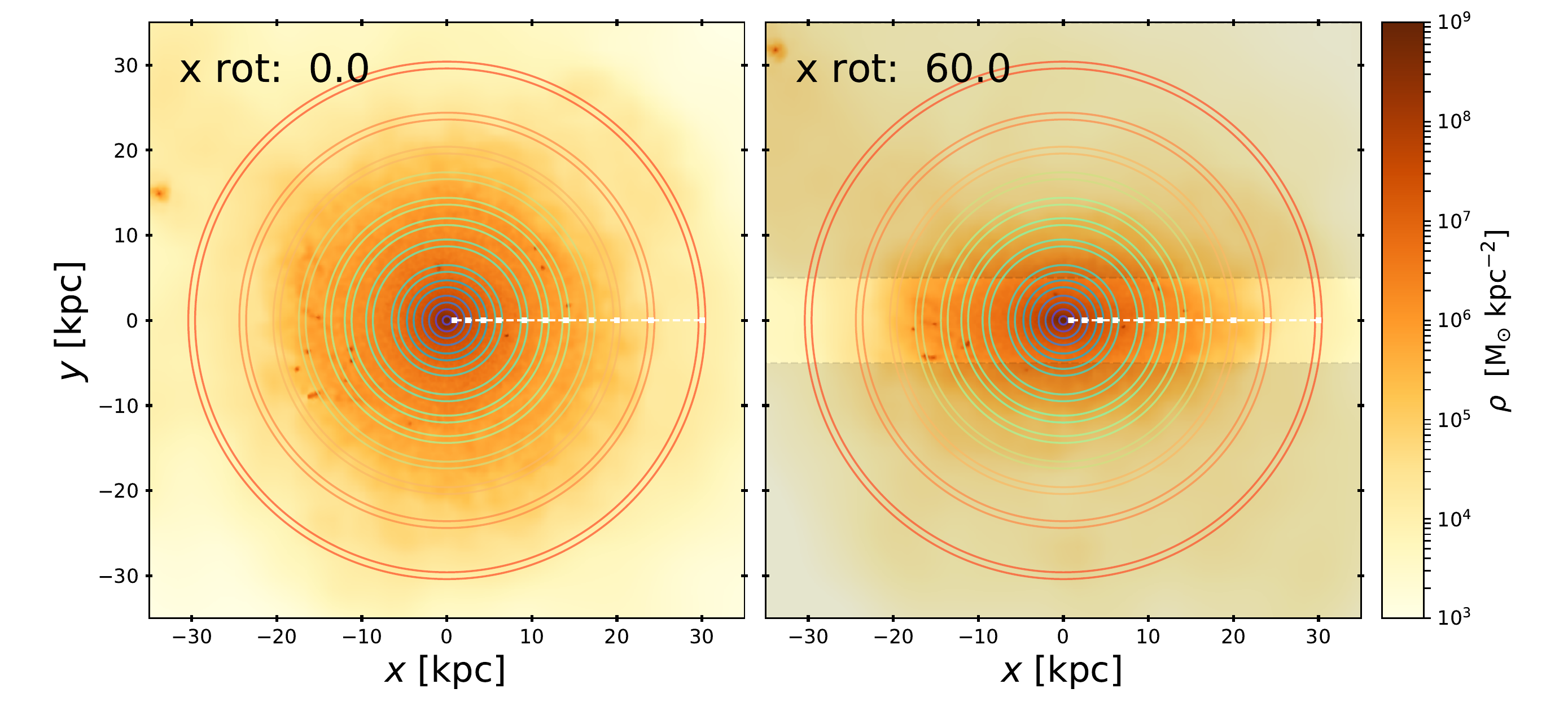}
    \caption{Concentric annuli spanning a region between 0.9 to 30 kpc from the galactic centre, chosen to select the star particles used to compute the SFH of our M33 candidate. The annuli are shown superimposed on the stellar mass density plot. The SFH at different radii will be computed for two configurations of the galaxy: face on, \textit{left panel}, and after a rotation of $60^{\circ}$,  \textit{right panel}. In this last case, we only considered the star particles within the annuli around the major axis of the galaxy, identified as a bright area in the right panel, i.e.  the region where the bulk stellar mass density resides, $|y|<5$ kpc.}
    \label{fig:sfh_annuli}
\end{figure}

In Fig. \ref{fig:sfh_annuli} we show the annular regions used to compute the SFHs \footnote{Considering that star particles migrate from their birth position, technically we calculate the Stellar Formation Time Distribution of the M33 look-alike at $z=0$. For clarity, we will maintain the SFH nomenclature throughout the paper.}, superimposed on the stellar mass density of our simulated galaxy, for the face-on view (left panel), and for the inclined view (right panel).
For this latter setup, instead of taking the star particles in the whole annular regions as done for the face-on analysis, we select the ones in the annular regions within the major axis ($|y|<5$~kpc) for this configuration. The area of the annuli that we consider in this case is indicated as a bright region in the right panel of   Fig. \ref{fig:sfh_annuli}. Once the annuli are defined, the star particles inside them are selected and  used to calculate the SFH at each radius\footnote{Each annular region is further subdivided in smaller regions, each subtending an arch of $0.8$ kpc across as we spam the 360$^{\circ}$ angle around each annuli, to mimick the size of the HST/ACS field of view. We perform the SFH analysis for such individual regions, as well as for the total amount of stellar particles within each annuli, in order to confirm that our results are not affected by a specific  position within each annuli.}.

\begin{figure*}
	\includegraphics[width=1\textwidth]{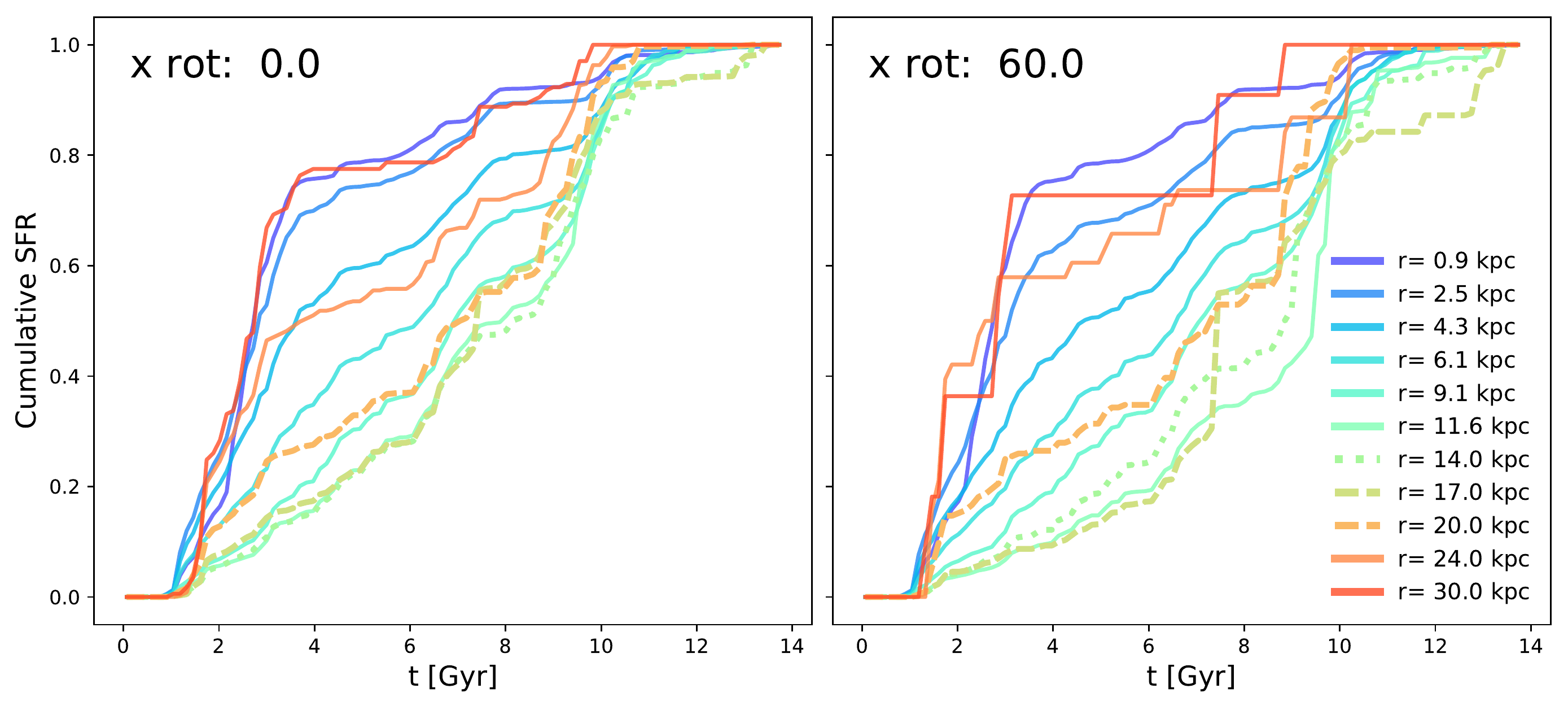}
    \caption{The cumulative normalized SFH of the simulated M33 galaxy, for several increasing radii out to a maximum radius of 30 kpc.
    \textit{Left}: SFH for the galaxy positioned face-on. \textit{Right}: SFH after a $60^{\circ}$ rotation around the $x$-axis of the galaxy. In both cases, moving  inside-out the median stellar age decreases, compatibly with an inside-out growth of the disc. However,  a reversal in the age gradient can be seen at large radii, where  an increasingly  higher  fraction of old stars contributes to the SFH. In both the face-on and the inclined case,  the radius at which the age reversal appears is between $r=17$ kpc and $r=20$ kpc (dashed lines). Projection effects may lead to a misidentification of the reversal radius, as seen from the median stellar age at $r=14$ kpc (dotted line).}
    \label{fig:sim_sfh}
\end{figure*}

In Fig. \ref{fig:sim_sfh} we show the results of the SFH analysis of the simulated M33, presented analogously to the observational data in Fig. \ref{fig:obs_sfh}, as cumulative SFHs at different radii, for the galaxy viewed face-on (left panel) and for the $60^{\circ}$ inclined view (right panel). Increasing radii are shown in different colors, from violet (0.9 kpc) to red (30 kpc).
Moving from the inner radii to the outer ones, the mean stellar age decreases, compatible with an inside-out growth of the disc.

However, at radii larger than $r=17$ kpc a reversal of the age gradient is observed in both the face-on and the inclined view,  just as found in  observations, with the percentage of old stars that contribute to the SFH rising again. The reversal region reported in W09+B11 ($r>9.1$ kpc) is smaller than the one found in our simulation: this should not be surprising given the different scale-lengths of the observed and simulated M33. In terms of relative disc scale lengths, the age reversal appears at a radius of about $5.1$ times the disc scale length in the observed M33, and $5.2$ times in the M33 look-alike. As mentioned in the previous section, we note that a break in the surface brightness and stellar surface density profiles of our M33 candidate appears at a radius $r=17.6$ kpc, coinciding with the radial region at which the age turnaround is found, which suggest that a correlation between the two effects could exist, similar to what has been discussed in W09+B11.

Comparing the inferred region of the reversal radius of the face-on and inclined views, we expect that projection effects in the inclined view would move the reversal radius closer to the galactic center after rotating the galaxy. The reversal region is instead found between $r=17$ and $r=20$ kpc (dashed lines) for both configurations. However, analyzing the surface brightness and stellar surface density profiles of the face-on view, we indeed find that the break moved to $r\sim19$ kpc (still within $17<r<20$ kpc). Thus, inspecting the cumulative SFH only provides a rough estimation of the turnaround radius' location, that is, it constrains the boundary of the region in which the turnaround radius is found. The age reversal detected in observations could be similarly affected by the  inclination of the galaxy itself and subsequent projection effects. Indeed, with a $60^{\circ}$  inclination with respect to the line-of-sight, the old stars in the outer region will overlap with the young stars in the disc, causing a higher contribution of old stars at smaller radii. This effect can be seen in the median formation time of the $r=14.0$ kpc region (dotted line), which shifted to a slightly older population after the inclination. 

We conclude that a reversal in stellar age gradient  appears for our M33 candidate with respect to the line-of-sight, and that the value of the reversal radius is  sensitive to the particular inclination chosen, being smaller for larger inclinations. While this change is not observed directly in the SFH of the face-on and inclined views of the galaxy, considering the correlation between the reversal radius and the break radius we are able to trace the projection effects through the break radius. Therefore, the deprojected reversal radius of the observed M33 may be at a different position than the one currently reported.
\subsection{Explanation of the Stellar Age Gradient} \label{sec:m33_age_grad.subsec:explanation}
As already touched upon before, there are several hypothesis that could explain the reversal of the stellar age gradient, such as mergers and stellar  radial migration.  To investigate which process determines the age turnaround in our simulation, we select the star particles that are within the M33 host at $z=0$ and trace them back in time to their birth redshift $z_{\rm birth}$ and position $r(z_{\rm birth})$, respectively.  

Then, by looking at their $r({z_{\rm birth})}/R_{\rm vir}^{M33}({z_{\rm birth})}$ distribution, we were able  to identify a main population of stars in the inner region of the galaxy, well separated from an outer stellar population found beyond  20 per cent of M33's virial radius.
Thus, we define \textit{in-situ} star particles as those  that  formed within  20 per cent of the virial radius of the main progenitor of M33 at the formation redshift, that is, whose birth radius satisfies the condition  $r({z_{\rm birth})}<0.2 \times R_{\rm vir}^{M33}({z_{\rm birth})}$. If, instead, the star particle has a birth radius larger than the above value it is defined as \textit{accreted}, since it is brought  into the main M33 galaxy via mergers and halos accretion. We performed some tests to check our selection criterion. We used a different classification of `in-situ' versus `accreted' stars by checking whether a star particle was born in the progenitor of M33, inside a subhalo, or outside the progenitor's virial radius. We also cleaned the disc from highly eccentric orbits that are not compatible with stars formed in the disc by making an eccentricity and vertical (z-component) velocity cut at $z=0$ of $e<0.6$ and $|v_{z}|<150$ km/s, respectively. We confirm that our tests had no influence on the results presented here and that the virial radius condition is  sufficient to ensure that the star particles are formed within the disc of the progenitor of M33.

\begin{figure}
	\includegraphics[width=\columnwidth]{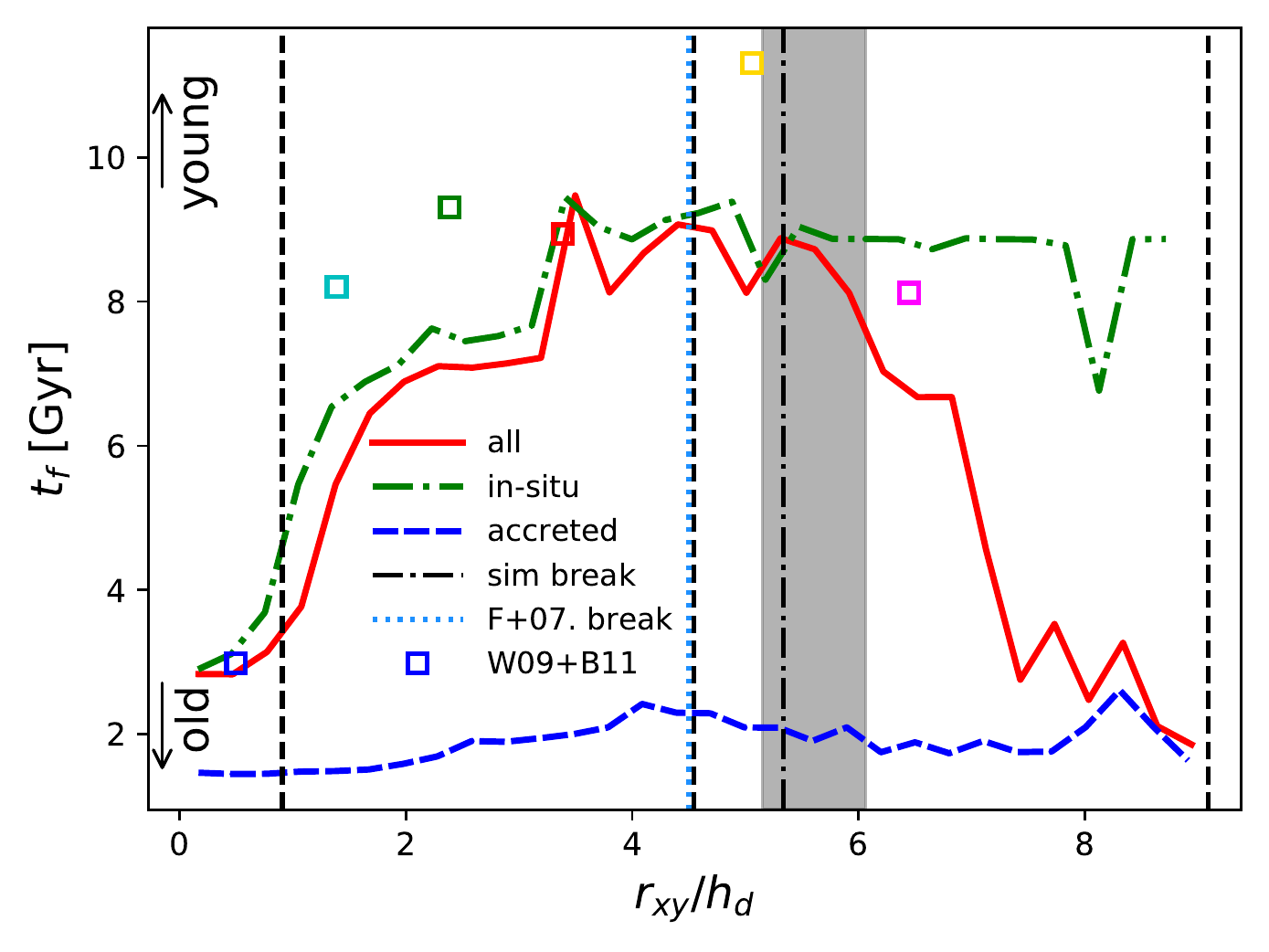}
    \caption{Projected (2D) median formation time of star particles in the inclined view, for increasing radial bins normalized by the disc scale length. The formation time profile of all star particles within M33 at $z=0$ is  shown as a red line, while in-situ and accreted stars are shown as green dash-dotted and blue dashed lines, respectively. Vertical dashed lines indicate the pre-turnaround (3-15 kpc) and the turnaround region (15-30 kpc) of the galaxy. The inferred median formation time for different radii from W09+B11 is shown as the color-coded markers. The blue vertical dotted line shows the surface brightness break radius from \citet{Ferguson07}, whereas the black dash-dotted vertical line shows the break radius in the M33 look-alike. The inferred age turnaround radius from the SFH of the M33 candidate in Fig. \ref{fig:sim_sfh} is represented by the gray shaded region. Note that the age turnaround disappears when only in-situ star particles are considered.}
    \label{fig:tform_2D_profile_30kpc}
\end{figure}

\begin{figure}
	\includegraphics[width=\columnwidth]{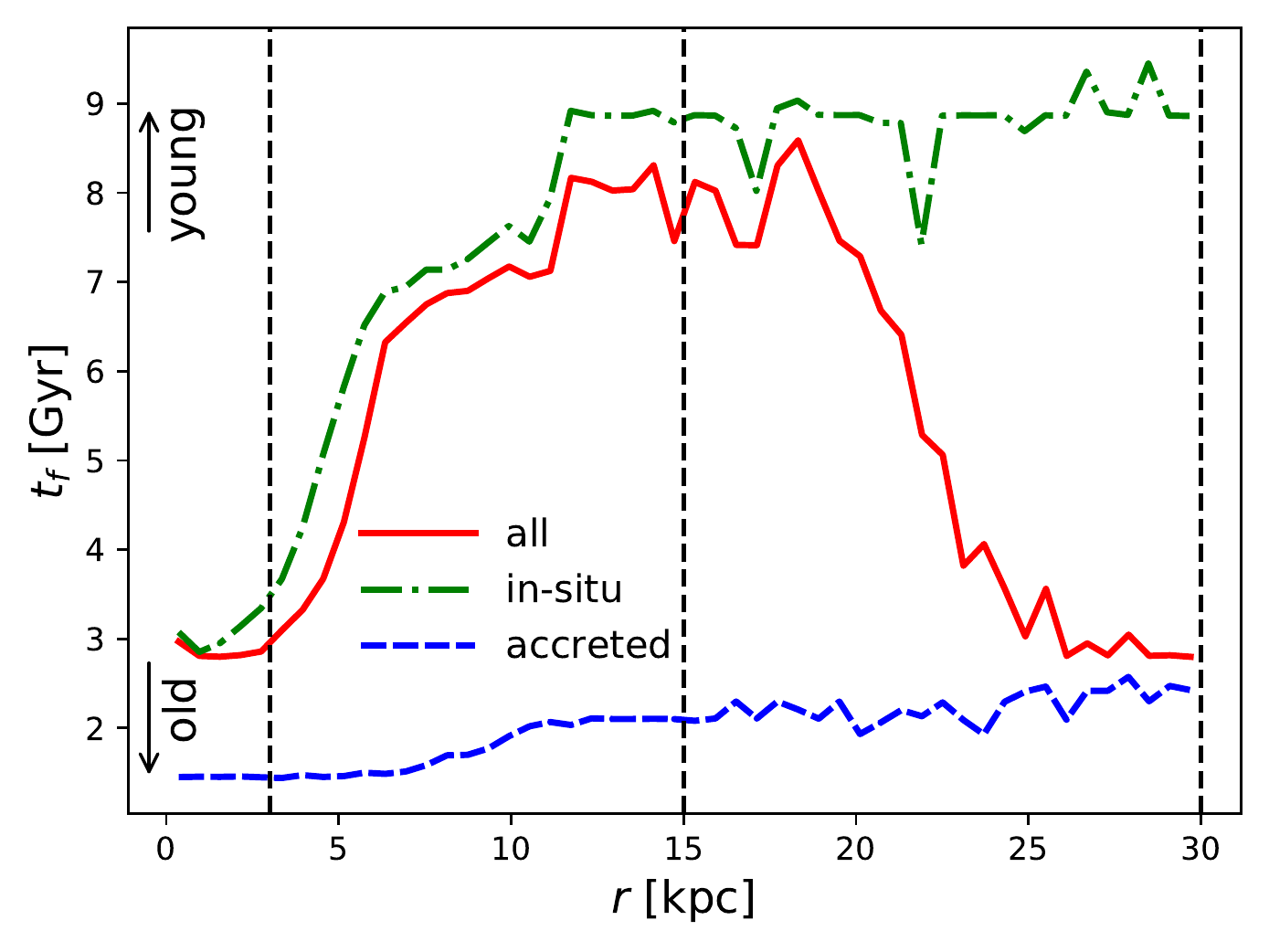}
    \caption{3D median formation time of star particles for increasing radial bins. We can see similar trends as the ones identified in the projected (2D) median formation time profile in Fig. \ref{fig:tform_2D_profile_30kpc}: a turnaround in the whole star population at $r>15$ kpc, a formation time flattening of the in-situ population, and an old flat distribution of star particles up to $r<30$ kpc. The fraction of in-situ stars within these regions is shown in Fig. \ref{fig:fraction_insitu}.}
    \label{fig:tform_profile_30kpc}
\end{figure}

In Fig. \ref{fig:tform_2D_profile_30kpc},  we show the projected (2D) median formation time for all (red solid line), in-situ (green dash-dotted line) and accreted (blue dashed line) star particles, as a function of radius (normalized by the disc scale length to correct for the size of the galaxies), for the inclined view of the M33 candidate. Along the simulated results we show the inferred median formation time profile from W09+B11 (color-coded markers), i.e. the formation time at which half of the cumulative SFR is reached at each radius. As we can see in the figure, the break radius of our simulation (dash-dotted black line) lies within the region of the turnaround radius inferred from the cumulative SFH (gray shaded region). Moreover, the maximum formation time (minimum age) of each profile is located at a similar relative projected radial positions, at $r_{\rm max,t_f}\sim4.5h_{d}\sim15$ kpc for the simulated profile, and at  $r^{M33}_{\rm max, t_{f}}\sim5h^{\rm M33}_{d}\sim9.1$ kpc for the observed one; and at a similar projected radius to their respective break radii, at $r_{\rm break}\sim5.3h_{d}\sim17.6$ kpc for the M33 candidate, and at  $r^{M33}_{\rm break}\sim4.5h^{\rm M33}_{d}\sim8.1$ kpc for the real M33. Once we scale the radial dependence of the profile to the corresponding disc scale length, the results are in good agreement with the observational trends. However, this has been done for a particular inclined initial configuration (i.e. our best configuration as introduced in Sec.~\ref{sec:m33_counterpart.subsec:validate_numerical_M33}). Although we obtain similar trends to the observational data, we are aware of the statistical limitation of the analysis. While the age turnaroud is evident when considering the full  sample of stars, with a median formation time decreasing from a maximum of $t_{f}\sim8$ Gyrs at $r_{\rm xy}\sim4.5h_d$ ($r_{\rm xy}=15$ kpc) to a minimum of $t_{f}\sim3$ Gyrs at $r_{\rm xy}\sim9.1h_d$ ($r_{\rm xy}=30$ kpc),  such a turnaround does not show up when only in-situ star particles are considered. Hence, the main driver of the reversal is not due to in-situ star particles, but rather to accreted ones. As we move further out from the galactic center, the in-situ stars  become progressively younger until reaching a radius of  $r_{xy}\sim4h_{d}$ ($r_{\rm xy}\sim13$ kpc), after which their age distribution flattens, with a median age of 4.7 Gyrs ($t_{f}\sim9$ Gyrs) for radii larger than 13 kpc. Note that, in an inside-out formation scenario we expect a monotonic increase of the formation time of the in-situ star particles with increasing radii, thus, the flattening of the profile clearly demonstrates that stellar migration also plays a role in the stellar age gradient of our simulated M33 (we will return to this point later). The same trends can be seen in the (3D) median formation time profile of the galaxy, as presented in Fig. \ref{fig:tform_profile_30kpc}: a turnaround in the whole star population at $r>15$ kpc, a formation time flattening of the in-situ population, and an old flat distribution of star particles up to $r<30$ kpc; with the only difference being the actual median formation time values of each component. In light of these results, in the following section we opted for a full 3D analysis of the origin of the turnaround.\\

\begin{figure}
	\includegraphics[width=\columnwidth]{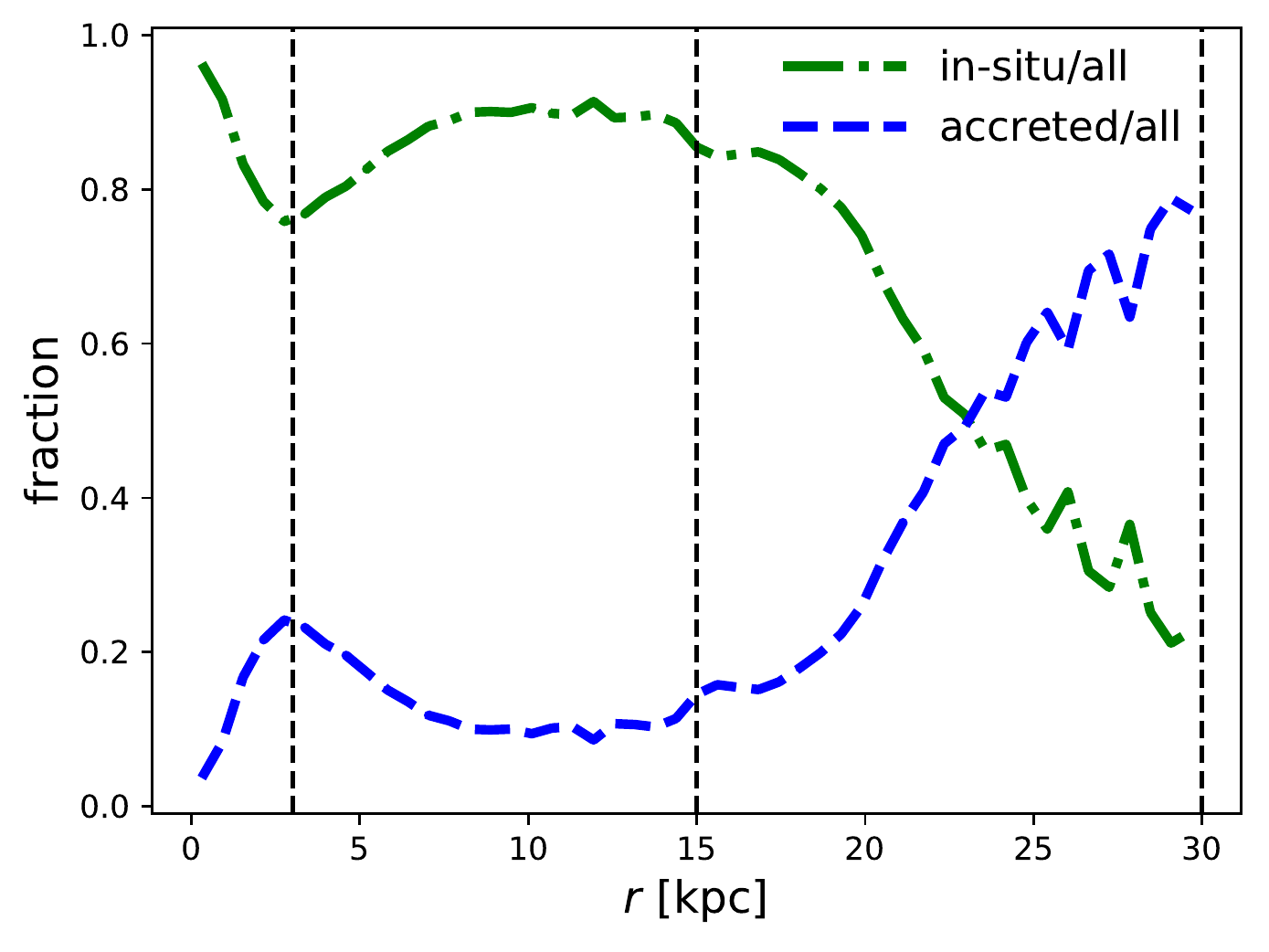}
    \caption{Fraction of in-situ and accreted stellar particles within radial bins, found in the simulated M33 at z=0. The in-situ stars made up   $\sim85$ per cent of the total stellar mass within $r=15$ kpc, while at larger radii their fraction decreases dramatically to about $\sim20$ per cent for $r=30$ kpc. Correspondingly, the fraction of  stars accreted via mergers increases with radius, causing the age-reversal observed at large radii. The peak of accreted stars seen in the inner radii, at $r=3-4$ kpc, is attributable to the last major merger that occurred in the early phases of the life of the galaxy, before $z=2$.}
    \label{fig:fraction_insitu}
\end{figure}

Star particles that get into the galaxy via mergers are the main drivers of the observed  age turnaround. The relative fraction of in-situ and accreted star particles within radial bins are shown in Fig. \ref{fig:fraction_insitu}: the in-situ star fraction within $3-15$ kpc is $\sim80$ per cent, while it decreases sharply as we move towards the outskirt of the galaxy, with a minimum of $20$ per cent star particles found at $r=30$ kpc being in-situ. The fraction of accreted stars, correspondingly, increases with radius: in the turn-around region, the accreted star fraction move from less than $20$ to almost $80$ per cent. We verified that the accreted stars that end up in the age reversal region (at $r>15$ kpc) are brought in through  minor mergers. Instead, the peak of accreted stars seen in the inner radii, at $r=3-4$ kpc, is attributable to the last major merger that occurred in the early phases of the life of the galaxy, before $z=2$.

\begin{figure}
	\includegraphics[width=\columnwidth]{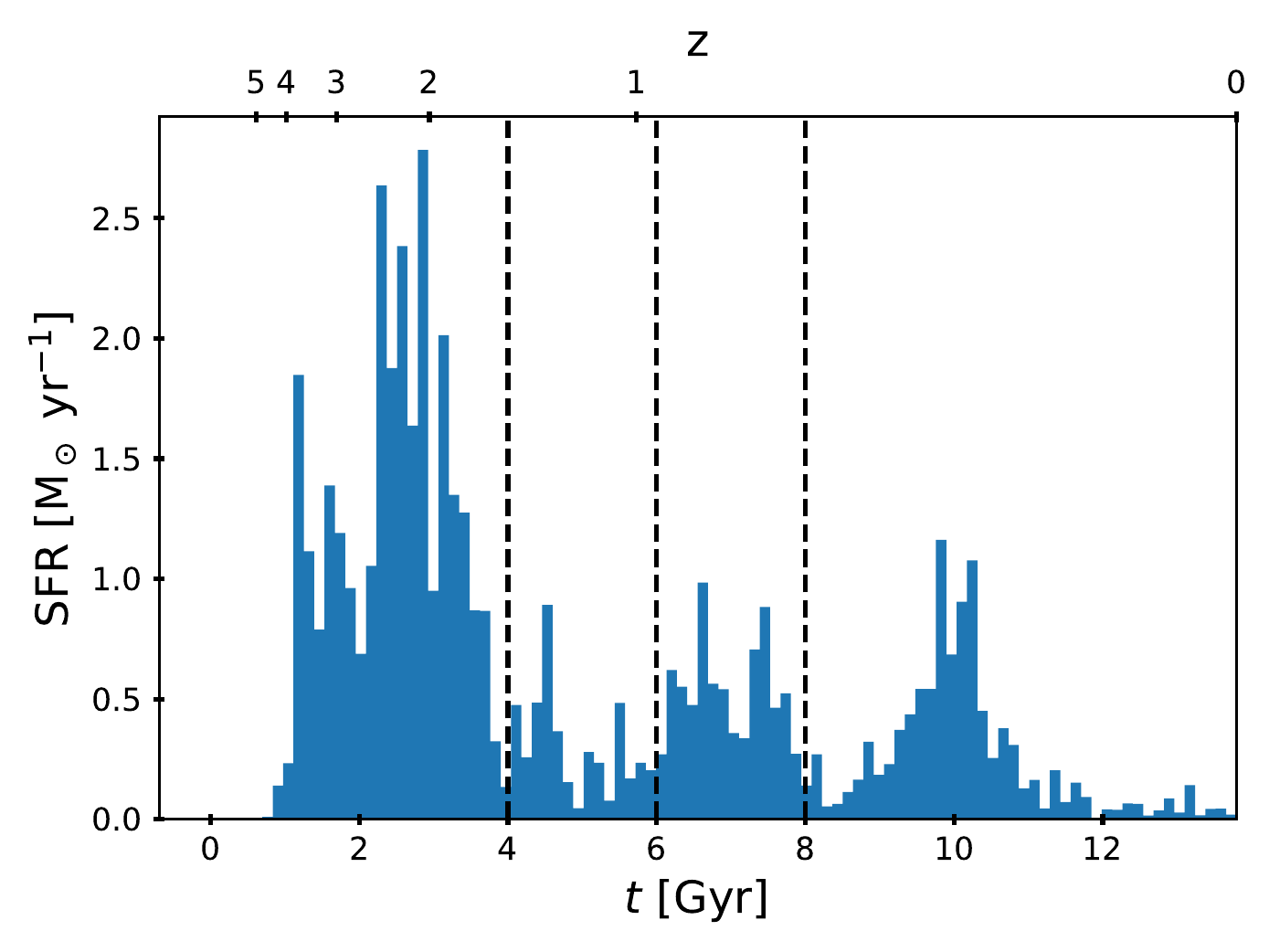}
    \caption{Integrated SFH of the simulated M33. We indicate the stellar age bins that will be later used for further analysis, selected following the main bursts of star formation.}
    \label{fig:sim_all_sfh}
\end{figure}

\begin{figure*}
	\includegraphics[width=\textwidth]{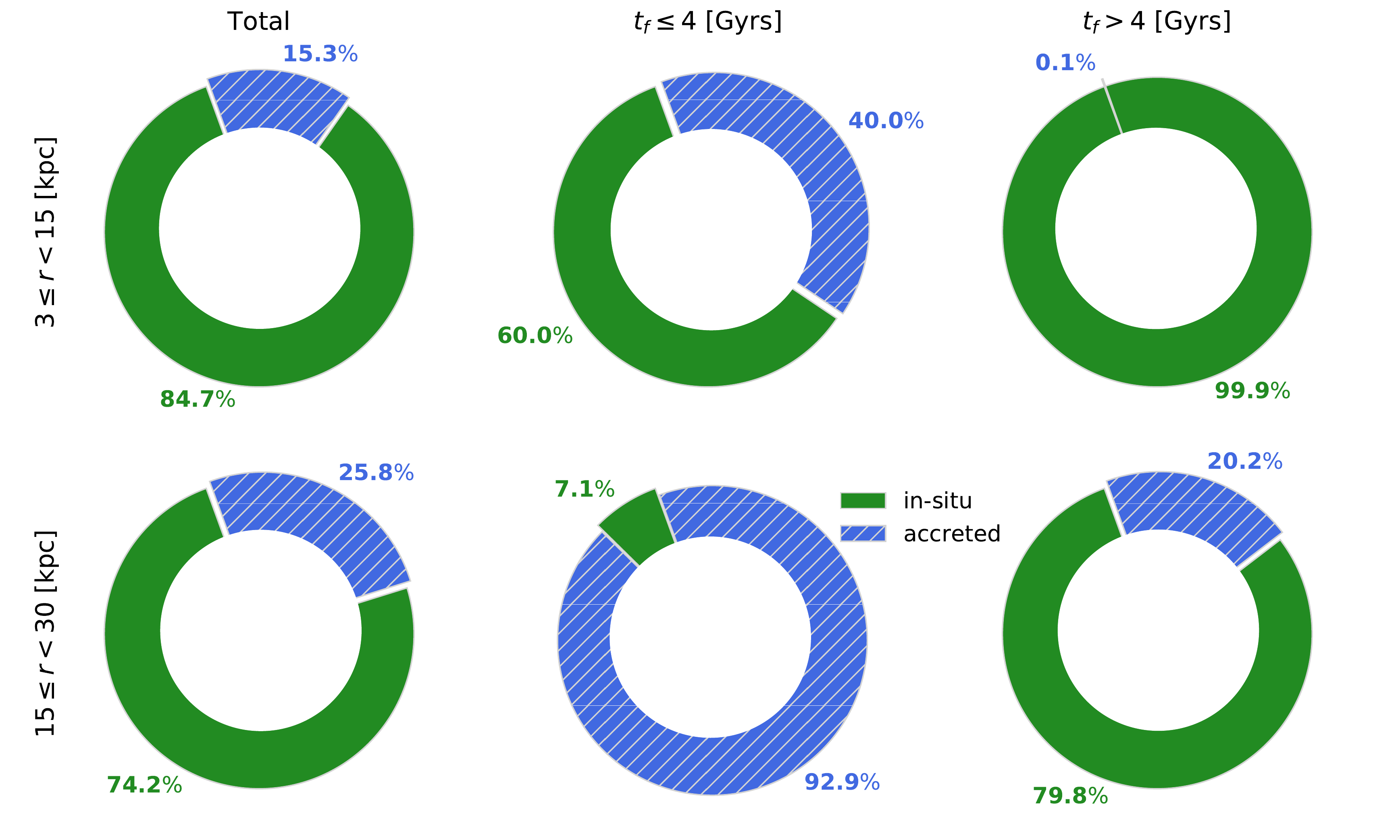}
    \caption{In-situ star particles percentages in the pre-turnaround region ($3<r\leq15$ kpc, top row) and in the turnaround region ($15<r\leq30$ kpc, bottom row). Only about $\sim7$ per cent of the old stars  ($t_{f}\leqslant$4 Gyrs) found in the outskirt of M33 (15 to 30 kpc), where the age-turn around is observed, are in-situ star particles.}
    \label{fig:in_situ_percent_pie}
\end{figure*}

To investigate the role of accreted stellar particles as a function of their age, we first aim at classifying the stars by their actual formation time. To this extent we present in Fig. \ref{fig:sim_all_sfh} the total, integrated SFH for the simulated M33 analogue at $z = 0$. We can observe several star forming bursts along its evolution. These bursts allow us to divide the stars into the following formation time bins: an old population of stars with $0 < t_{f} \leq 4$ Gyr, a old-to-intermediate stellar population corresponding to a small burst with $4 < t_{f} \leq 6$ Gyr, an intermediate-to-young stellar population with $6 < t_{f} \leq 8$ Gyr, and the rest of the stellar population with $t_{f} > 8$ Gyr corresponding to the last major stellar burst.
We then show, in Fig. \ref{fig:in_situ_percent_pie}, the percentage of in-situ star particles in the pre-turnaround region, $3<r\leq15$ kpc (top row),  and in the turnaround area $15<r\leq30$ kpc (bottom row)\footnote{Since we are interested in disc-stars, we avoid the central bulge region of our simulated galaxy when computing percentages, and select  star particles  with $r>3$ kpc.}. From left to right, we indicate the full percentage of in-situ stars, and the relative percentages of in-situ stars once  binning by formation time $t_{f}$. In-situ star particles  account for $\sim85$ per cent of the total stellar mass  in the inner region of M33, and only for $\sim74$ per cent of the total star mass in the outer region. Therefore, the total accreted star particle fraction increases from $\sim15$ to $\sim26$ per cent  when moving towards the outskirts of the galaxy. As expected for an inside-out growth scenario, almost all the stars with a formation time $t_{f}>4$ Gyrs are born in-situ, at any radial region.
While, strikingly, only about $\sim7$ per cent of the old stars  ($t_{f}\leqslant$4 Gyrs) found in the outskirt of M33 (15 to 30 kpc), where the age turnaround is observed, are in-situ star particles. In the radial region corresponding to the observed age reversal,  $\sim93$ per cent of the stars with old formation time are accreted via mergers, being the cause for  the turnaround in the formation time profile. 

\begin{figure*}
	\includegraphics[width=\textwidth]{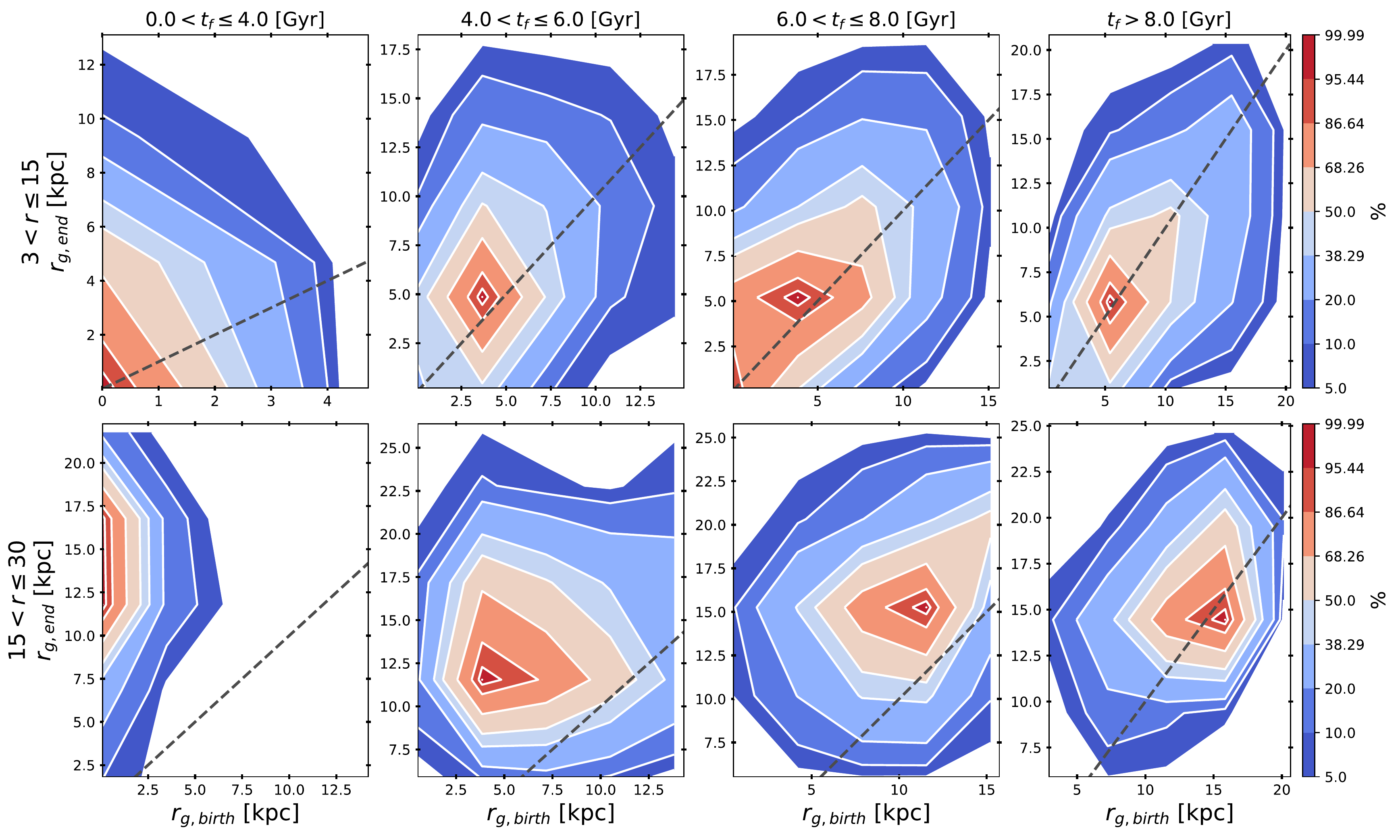}
    \caption{Stellar migration density probability, end guiding radius $r_{g,\rm end}$ versus birth guiding radius $r_{g,\rm birth}$ for in-situ star particles. \textit{Top row:} stellar migration in the inner region, $3<r\leq15$ kpc. \textit{Bottom row:} stellar migration in the outer region, $15<r\leq30$ kpc. The dashed line represents the locus of star particles for which  no migration occurred. Old ($t_{f}\leq 4$ Gyr) in-situ stars found at large radii (r$\geq15$ kpc) have migrated from the inner region towards the outskirts of M33.}
    \label{fig:stellar_migration}
\end{figure*}

Next, to verify whether stellar migration  also plays a role in the age reversal, we quantify the amount of radial migration that has taken place in the disc. To this extent, we need to estimate the change in angular momentum, which is equivalent to the change in guiding radius $(r_g)$ since a star particle's formation time. We approximate $r_g$ for each star particle as described in \citet{Minchev14} by calculating the birth and end guiding radii as:

\begin{equation*}
r_{g}=\frac{L}{v_{\rm circ}(r)} = \frac{r\cdot v_{\phi}}{v_{\rm circ}(r)} 
\label{eq:guiding_radius}
\end{equation*}\\

\noindent where $L$ and $v_{\phi}$ are the angular momentum and rotational velocity of the star particle, respectively, and $v_{\rm circ}$ is the rotation curve. The guiding radius takes into account stars with high eccentricity orbits by comparing the star's tangential circular velocity, $v_{\phi}$, with the circular velocity of the galaxy at the star's radius, $v_{\rm circ}(r)$. This allows us to know the change in the angular momentum of the star particles, which is equivalent to the change in the guiding radius at $z=0$ $(r_{g, \rm end})$, and the guiding radius at the birth redshift of the star particle $(r_{g,\rm birth}$). By using the guiding radius we avoid mistaking stars with high eccentricity orbits as instances of radial migration. 

We now present in Fig. \ref{fig:stellar_migration} the guiding radius at $z=0$ ($r_{g,\rm end}$) vs. the guiding radius at the birth redshift ($r_{g,\rm birth}$) for the star particles that were born in M33's main progenitor, i.e. in-situ ones, selected from two regions in the formation time profile, $3<r\leq15$ kpc (top row) and $15<r\leq30$ kpc (bottom row), further separating the stars using aforementioned formation time bins (from left to right, we move from old to young stars). Black dashed lines indicate the locus of stars that have not changed their position, i.e. whose guiding birth radius $r_{g,\rm birth}$ is the same as their end one $r_{g,\rm end}$. In the inner region, most of the old, $t_{f}\leq 4$ Gyr, population moved to slightly greater  radii ($\sim 70$ per cent  of stars moved from $r_{g,\rm birth}\sim0.5-1.5$ kpc to $r_{g,\rm end}\sim2-4$ kpc), whereas almost all the younger stars ($t_{f}>4$ Gyr) remained close to where they were born, i.e. close to the black dashed line. Thus, radial migration is observed to some extent for the old population of in-situ stars in the inner region of the galaxy:  their final guiding radius, however, it is still smaller than the radius at which the age turnaround is found.\footnote{Note that we initially selected the star particles by their radius $3<r\leq15$ kpc, and by using the guiding radius $r_{g}$ we find stars with $r_{\rm g,end}<3$ kpc, outside the originally selected region at $z=0$. This indicates that in this region there are some star particles that are on highly eccentric orbits. Although we find them at a particular radius at $z=0$, their guiding radius reveals they have different average radii.} In the outer region, however, we see a considerable radial migration  for the old population ($t_{f}<4$ Gyr), most of which migrate from $r_{\rm g,birth}\sim0$ kpc to $r_{\rm g,end}\sim12-17$ kpc, and few of them reaching as far out as $r_{\rm g,end}\sim20$ kpc. The rest of the stellar populations present less and less radial migration the younger they are. Thus, the old in-situ stars, which made up only the $7$ per cent of the old stars found in the outer region of M33, have undergone thorough radial migration, i.e. they are stars that were born close to the galaxy's disc plane center and have migrated outwards. 
However, given their small number compared to the fraction of old accreted stars, migration of in-situ stars  alone is not sufficient to explain the age turnaround found in Fig. \ref{fig:tform_profile_30kpc}: without accretion we could not observe this age reversal. Hence, we conclude that the main driver behind the reversal of the age gradient is stellar accretion and, to a lesser extent, stellar migration.

\subsection{Observational predictions} \label{sec:m33_age_grad.subsec:observation_predict}
In order to verify that the age turnaround is mainly driven by stellar accretion, one would need to be able to differentiate \--- observationally \---  accreted from in-situ stars.
Since both groups of  stars share the same location in the galaxy at $z=0$, we must rely on their kinematics for this aim.
In this section, we investigate the kinematic properties of the two populations of star particles found in the outer region  of  our simulated Triangulum galaxy.

\begin{figure}
	\includegraphics[width=\columnwidth]
{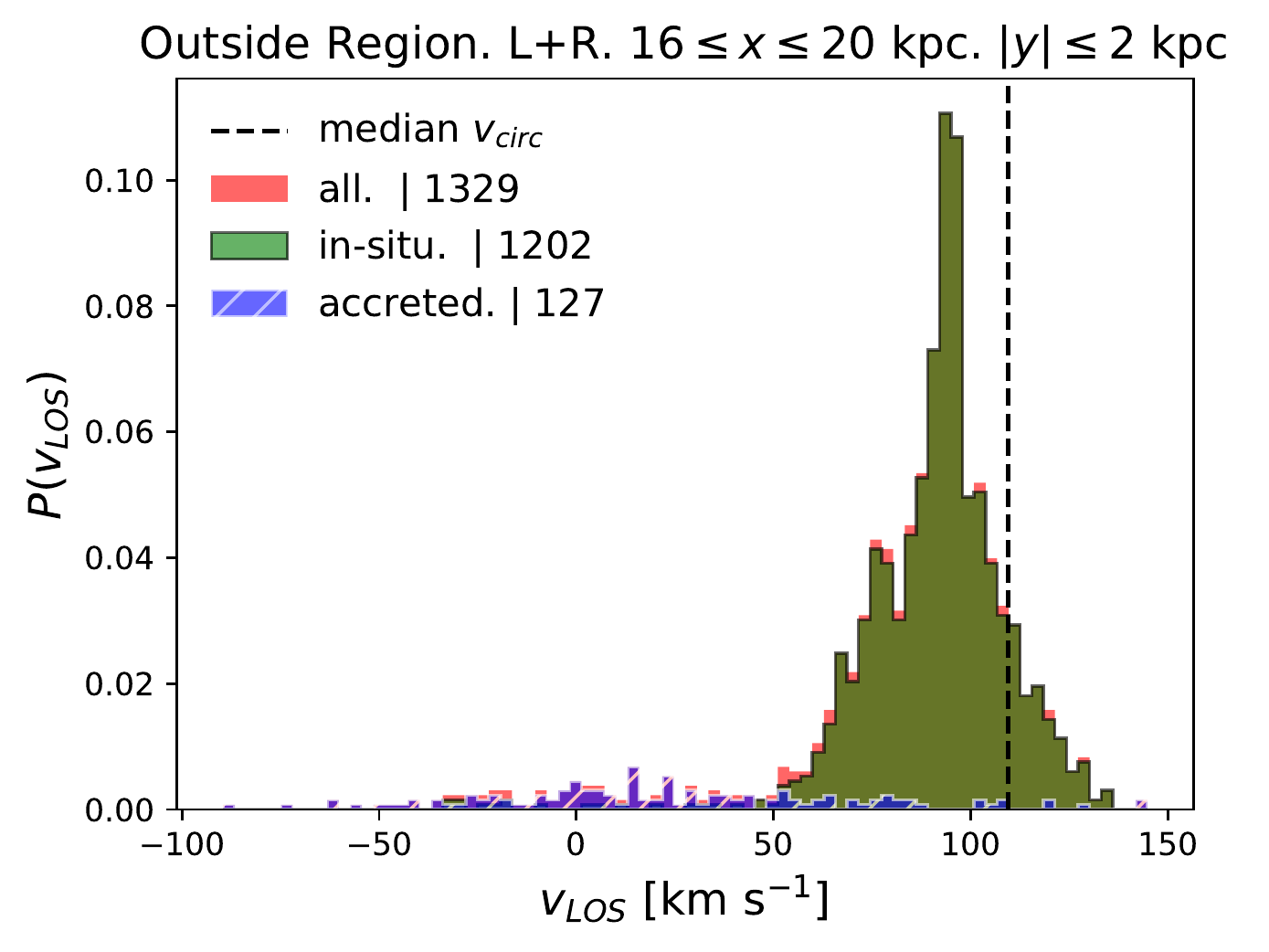}\caption{Line-of-sight velocity histograms of all (red), in-situ (green) and accreted (blue hatched fill) star particles, with the number of star particles in each category, within a 4$\times$4 kpc$^2$ region centered at the turnaround radius, after inclining the galaxy 60$^{\circ}$, as the observed M33. The vertical dashed line is the galaxy's circular velocity corrected for inclination. In-situ star particles co-rotate with the galactic disc, and lag with respect to the rotation curve due to the asymmetric drift effect, resulting from the non-zero stellar velocity dispersion. Accreted ones have a random distribution of their $V_{\rm los}$, with a large dispersion.}
    \label{fig:dispersion}
\end{figure}

Using the same (best) inclined configuration of our numerical M33, we study the line of sight velocity $V_{\rm los}$  of the star particles in the disc. We selected the projected star particles within a square region of 4 kpc in side, alongside the major axis of M33, centered on the age turnaround region, $16\leqslant x \leqslant20$ kpc  and $|y|\leqslant 2$ kpc. To improve the sample of stars, and to avoid any preferential direction on the mayor axis, we performed the selection for both the right-hand side (x>0) and the left-hand side (x<0) of the galaxy, and combined the selected stars taking into account the sign change due to the rotation of the star particles in the disc. 
We then plotted a histogram of the star particles' line of sight velocity $V_{\rm los}$ (in projection, identified with the z axis-component of the star particle's velocity), in Fig. \ref{fig:dispersion}. The total, in-situ and accreted stars, are shown as red, green and blue histograms, respectively, along the number of stars in each category. The vertical dashed line at $V_{\rm los}\sim 109$ km/s is the galaxy's circular velocity corrected for inclination.
The difference between the accreted star particles and the in-situ ones is clear: while the latter are co-rotating with the disc, with a peak rotation  of $V_{\rm los}\sim 94$ km/s  and a relatively small velocity dispersion, $\sigma \sim 19$ km/s, the accreted stars have randomly distributed $V_{\rm los}$, averaging to 0, i.e. implying radial orbits in the absence of angular momentum, and displaying a much higher dispersion, $\sigma \sim 46$ km/s.

Future surveys will potentially be able to identify rotationally supported stars in the outskirts of M33, a signature that they are in-situ stars, and differentiate them from accreted ones.
Our model predicts that the in-situ stellar population in the outskirsts of the galaxy should have a flat radial age gradient as a result of accretion and stellar migration: in our particular simulation, the median stellar age of in-situ, co-rotating stars at large radii is $4-5$ Gyrs.
Accreted stars, on the contrary, are expected to be uniformly old, with a median age of $\sim$11 Gyrs.

\section{Conclusions} \label{sec:conclusions}
We presented properties of a M33-analogue galaxy, simulated within the framework of Constrained Local UniversE Simulations \citep{Gottloeber10,DiCintio12,Carlesi16},  run with the code \code{GASOLINE} \citep{gasoline} and including supernova feedback  \`{a} la \cite{Stinson06}, which allows for an efficient regulation of star formation within galaxies.
 
The properties of the simulated M33 are in fair agreement with observational data from \citet{Ferguson07,Corbelli14} and \citet{Kam17}, in terms of mass, rotation velocity, and surface brightness. Our simulated M33 has a total virial mass of $2.7\cdot10^{11}$M$_{\odot}$ with a stellar mass of $5.1\cdot10^{9}$M$_{\odot}$, placing the galaxy on the correct expectations from abundance matching prediction, a thin extended disc with scale length value of $\sim 3.2$ kpc, and a rotation curve whose maximum value of  $V=127.6$ km/s is reached at a radius of $r\sim17$ kpc, similar to what is reported in observations.
The M33 simulated candidate does, however,  have a small bulge in its inner region, not observed in the Triangulum galaxy: consequently we have avoided the  region of the galaxy $r<3$ kpc in our analysis.
In the CLUES-M33 analogue, we observe a trend of decreasing stellar age as we move towards outer radii, a sign of an inside-out formation of the disc (e.g. \citealt{Pilkington12}), in which old stars are found within the inner most regions of a galaxy and young stars in the outskirt of the disc, as a result of newly accreting gas.
In order to compare the observational results with our simulations we selected stellar particles within concentric annuli regions  of similar thickness  as the field of view of the HST/ACS camera, that has been used to derive the cumulative star formation history (SHF) of M33 along its major axis \citep{Williams09,Barker11}.

Interestingly, similarly to what was found in observations, the age gradient of stars in the simulated M33  shows a turnaround at large radii, r$\geq$ 17-20 kpc, with the percentage of old stars ($t_{f}<4$ Gys) increasing from $\sim20$ to $\sim80$ per cent, moving from radii 14 to 30 kpc. Several proposals appear in the literature for explaining such age profile turn-around: stellar migration \citep{Roskar08a,Roskar08b,RuizLara16a}, projection effects \citep{Barker11}, a decrease in gas volume density possibly related to the warps in the disc \citep{SanchezBlazquez09}, and accretion events from mergers \citep{Gill05,Sales2007,Brook12,RuizLara16a}.

In this work, we demonstrate that this age reversal is mostly a result of accretion of old ($t_{f}\leq4$ Gyrs) stars  from merging satellite galaxies into the main host galaxy and, to a much lesser extent, of stellar migration of  old in-situ stars from the central regions towards the outskirts of M33.  This result is in agreement with previous work by \cite{RuizLara16a}, using more massive galaxies. Indeed, at large radii where the age turnaround is found, about $93$ per cent of the old stars come from accretion events, while only a mere $7$ per cent were formed within the M33 galaxy disc (i.e. are in-situ): the reversal in the stellar age gradient disappears when considering only in-situ star particles. This suggests that accretion from mergers are the origin of the turnaround in our simulated M33. 

This scenario could be verified observationally,  studying the kinematic of stars in the outer fields of M33: in-situ stars should be co-rotating with the galactic disc, and should have a relatively small velocity dispersion $\sigma$, while accreted stars, which are kinematically hot, are expected to have a random distribution in their line-of-sight velocity, and to show a large velocity dispersions (in our model, more than two times higher than the $\sigma$ of in-situ stars at the same radius). Moreover, the median age of the rotationally supported, in-situ stars, should indicate that this stellar population is young (median age of 4-5 Gyrs), unlike the pressure supported, accreted stars, causing the age turnaround, which should  be all old (median age $\sim$ 11 Gyrs).

We highlight that the method used is sensitive to projection effects. While changing the inclination of the galaxy did not induce an apparent change in the turnaround radius region observed in the cumulative star formation history of the galaxy (i.e. $17\leq r\leq20$ kpc), since it only provides a rough estimation of the turnaround radius; considering the correlation between the age reversal radius in the median formation time profile and the break radius of the surface brightness and stellar mass density profiles we are able to trace the projection effects through the break radius. Therefore, projection effects must be thought of carefully since they might play an important role in the determination of the true age turnaround radius.

Similar to what was found in observations \citep{Ferguson07}, a break in the surface brightness profile of our M33 candidate in its inclined configuration appears at $r=17.6$ kpc ($5.3$ times its disc scale length), coinciding with the radius at which the age turnaround is found. Moreover, following the \citet{Martin-Navarro12,Martin-Navarro14} classification, we detect a truncation coexisting with an up-bending of the surface brightness profile associated with the stellar halo component of the simulated galaxy at $r=25$ kpc. Similar results are obtained from the stellar surface mass density profile of the M33 candidate, i.e. a comparable disc scale length, and a break and a truncation at the same radii. Thus, both the radial mass distribution of the star particles and their age/metallicity contributes to the reversal of the age gradient at the outskirts of the galaxy. Recently, \citet{trl17}  showed, using simulations, that breaks are a consequence of the combined effects of outward-moving and accreted stars, in good agreement with our results. 

Finally, we note that \citet{RuizLara16a} found similar results when studying Milky Way-mass galaxies in the RADES \citep[\code{RAMSES} \textit{Disc Environment Study} simulations,][]{Few12}. In those simulations, the age reversal appears due to a combination of an inside-out growth of the disc, stellar migration (both inwards and outwards) of disc stars and accretion from old satellites: interestingly, as in our model, the age reversal  was  still recovered after suppressing stellar radial motion, indicating the minor relevance of stellar migration in  generating the age upturn observed at large radii in massive galaxies.

In the future we intend to verify if the accretion phenomenon causing the age turnaround is dependent on the specific  mass accretion history of each galaxy: in order to shed light on this we would need a large statistical sample of hydrodynamically simulated halos of M33's mass. The recently developed Local Group Factory \citep{Carlesi16} could be used to this aim.

\section*{Acknowledgements}
RM, AK, and CB are supported by the {\it Ministerio de Econom\'ia y Competitividad} and the {\it Fondo Europeo de Desarrollo Regional} (MINECO/FEDER, UE) in Spain through grant AYA2015-63810-P. ADC acknowledges financial support from the Karl-Schwarzschild fellowship  program as well as a Marie-Sk\l{}odowska-Curie Individual Fellowship grant, H2020-MSCA-IF-2016 Grant agreement 748213 DIGESTIVO. AK is also supported by the Spanish Red Consolider MultiDark FPA2017-90566-REDC and further thanks Jay-Jay Johanson for the long term physical effects.  CB further thanks the MICINN (Spain) for the financial support through the Ramon y Cajal programme.  IM acknowledges support by the Deutsche Forschungsgemeinschaft under the grant MI 2009/1-1.

We would like to thank the anonymous referee for their very detailed and constructive comments that helped to improve the paper. We further thank Patricia S{\'a}nchez-Bl{\'a}zquez for valuable discussions.

This research has made use of NASA's Astrophysics Data System (ADS) and the arXiv preprint server.



\bibliographystyle{mnras}
\bibliography{archive}



\appendix

\section{Determination of the disc scale length of the M33 candidate}\label{app:disc_scale_length}
To obtain luminosities from the simulation we used the routines available from \code{PYNBODY}. The code uses the Padova simple stellar populations (SSPs) isochrones and evolutionary tracks \citep{Marigo08,Girardi10}, with no dust extinction, to create a table which returns, for a set of ages and metallicities, a magnitude in the desired photometric system (see \url{http://stev.oapd.inaf.it/cgi-bin/cmd} for an overview of the different settings available for the table). \code{PYNBODY} reads the star particle's ages and metallicities returned from the simulation and interpolates the aforementioned table to associate magnitudes/luminosities to each star particle, and consequently, to the M33 candidate. 

The $i-$band surface brightness profile has been fitted to a Sersic bulge and two exponential discs components corresponding to the inner and outer discs. Consequently, we used the following intensity profiles:
\begin{align*}\label{eq:surfb_intensity_fit}
I_{\rm bulge} (r)   &= B_{e}\exp{\left\lbrace-b_{n}\left[(r/r_{e})^{1/n} -1\right]\right\rbrace}\\
I_{\rm disc,in} (r) &= D_{\rm 0,in}\exp{\left\lbrace -(r/h_{\rm d,in})\right\rbrace} \numberthis\\
I_{\rm disc,out} (r) &= D_{\rm 0,out}\exp{\left\lbrace -(r/h_{\rm d,out})\right\rbrace}
\end{align*}
where $n$ is the Sersic index, $b_{n}\approx1.9992n - 0.3271$ (\citealt{Capaccioli89} approximation), $r_e$ the effective radius, $h_{d}$ the disc scale length, and $B_{e},D_{0}$ the intensity at the effective radius and at $r=0$, respectively. Thus, for the $i-$band, the surface brightness profile is:
\begin{align}\label{eq:surfb_formula}
\mu_{i}(r)=25.73-2.5\log_{10}{I_{i}(r)} \textrm{ mag arcs}^{-2}
\end{align}
where $I(r)$ is expressed in L$_\odot$ pc$^{-2}$.

The fitting has been done in the following regions: the bulge, $r\leq2$ kpc; inner disc, $5\leq r \leq14$ kpc; and outer disc, $16.5\leq r \leq23$ kpc. These regions were chosen in order to avoid radial ranges where we see an overlap of galactic components, and to minimize contamination from old stars at the outskirts of the galaxy $r\sim 30$ kpc. The same regions were used for the fit of the face-on view. The best-fit values for the inclined configuration are shown in Tab.\ref{tab:best_fit_surfb}.

To obtain the y-axis normalization value for the simulated profiles we used the inner disc fit models and best-fit parameters from Tab. \ref{tab:best_fit_surfb}, and linearly interpolated the curves at the desired radial value $r=h_{\rm d}$. This has been done in order to avoid the contribution of the bulge when determining the normalization. On the other hand, since the observed M33 has no bulge component, we linearly interpolated the observational values directly at the (observational) disc scale length $h^{\rm M33}_d=1.8$ kpc.


\begin{table}
\centering
\caption{Best-fit values for the surface brightness profile of the simulated M33 in the inclined configuration.
}
\label{tab:best_fit_surfb} 
\begin{tabular}{cccc}
\hline
\hline
M33       & $\mu_e(\mu_0)$[mag arcs$^{-2}$] & $r_e(h_d)$[kpc] & $n$\\
\hline
Bulge     & $21.2\pm0.2$                    & $1.3\pm0.1$     &  $1.3\pm0.2$ \\ 
\hline
Inner disc& $22.9\pm0.2$                    & $3.3\pm0.1$     & - \\
\hline
Outer disc& $17.1\pm0.7$                    & $1.68\pm0.09$   & - \\ 
\hline
\hline
\end{tabular}
\end{table}


\bsp	
\label{lastpage}
\end{document}